\documentclass[onecolumn,amsmath,amssymb,12pt,superscriptaddress,nofootinbib]{revtex4}
\usepackage[bookmarks,linktocpage, colorlinks=true, plainpages = false, citecolor = blue,  linkcolor=blue, urlcolor = blue,
filecolor = blue]{hyperref}

\usepackage{graphicx}
\usepackage{dcolumn}
\usepackage{bm}
\usepackage{braket}

\input epsf
\def\gsim{ \lower .75ex \hbox{$\sim$} \llap{\raise .27ex \hbox{$>$}} }
\def\lsim{ \lower .75ex \hbox{$\sim$} \llap{\raise .27ex \hbox{$<$}} }

\newcommand{\be}{\begin{equation}}
\newcommand{\ee}{\end{equation}}
\newcommand{\bea}{\begin{eqnarray}}
\newcommand{\eea}{\end{eqnarray}}

\def\a{\alpha}

\def\s{\sigma}

\def\e{\epsilon}

\def\k{\kappa}

\def\th{\theta}

\def\s{\sigma}

\def\pt{\partial}

\newcommand{\smallWidthLeft}{239pt}
\newcommand{\smallWidthRight}{229pt}

\def\bea{\begin{eqnarray}}
\def\eea{\end{eqnarray}}

\numberwithin{equation}{section}

\renewcommand{\a}{\alpha}
\renewcommand{\k}{\kappa}

\renewcommand{\th}{\theta}

\begin{document}
\allowdisplaybreaks
\begin{titlepage}

\title{On the Quantum--To--Classical Transition for Ekpyrotic Perturbations}

\author{Lorenzo Battarra}
\email[]{lorenzo.battarra@aei.mpg.de}
\author{Jean-Luc Lehners}
\email[]{jlehners@aei.mpg.de}

\affiliation{Max--Planck--Institute for Gravitational Physics (Albert--Einstein--Institute), 14476 Potsdam, Germany}

\begin{abstract}

\vspace{.3in}
\noindent We examine the processes of quantum squeezing and decoherence of density perturbations produced during a slowly contracting ekpyrotic phase in which entropic perturbations are converted to curvature perturbations before the bounce to an expanding phase. During the generation phase, the entropic fluctuations evolve into a highly squeezed quantum state, analogous to the evolution of inflationary perturbations. Subsequently, during the conversion phase, quantum coherence is lost very efficiently due to the interactions of entropy and adiabatic modes. Moreover, while decoherence occurs, the adiabatic curvature perturbations inherit their semi-classicality from the entropic perturbations. Our results confirm that, just as for inflation, an ekpyrotic phase can generate nearly scale-invariant curvature perturbations which may be treated as a statistical ensemble of classical density perturbations, in agreement with observations of the cosmic background radiation.

\end{abstract}
\maketitle

%\tableofcontents

\end{titlepage}

\section{Introduction}

Cosmological observations, most recently by the Planck satellite \cite{Ade:2013lta}, indicate conclusively that at early times our universe was exceptionally flat on average, but, crucially, also contained a near-Gaussian distribution of small and nearly scale-invariant density perturbations. These small (classical) perturbations are thought to provide the seeds for the formation of the whole cosmic web of galaxies via subsequent gravitational collapse \cite{Frenk:2012ph}. An open question is still what the origin of these fluctuations was.

The most popular explanation is the theory of inflation \cite{Guth:1980zm,Linde:1981mu,Albrecht:1982wi}, which proposes that the primordial density perturbations arose from the amplification of quantum fluctuations of a scalar field during a phase of accelerated expansion of the universe. The transition from quantum fluctuations to a statistical distribution of classical density perturbations has been studied in detail in this context (see for example \cite{Guth:1982ec,Brandenberger:1990bx,Albrecht:1992kf,Polarski:1995jg,Prokopec:2006fc,Kiefer:2007zza,Martin:2012ua}), and it was found that there are two effects that support the classical interpretation of the resulting fluctuations: the first is that the (approximately Gaussian) quantum state evolves into a highly squeezed state, which means that in the field amplitude/momentum plane there is one direction along which the quantum mechanical uncertainty is very small, and a perpendicular direction in which it is very high. Such a state is far from classical, but can in fact be reinterpreted as a statistical distribution of classical states if in addition the density matrix describing these quantum fluctuations is approximately diagonal. The density matrix must be approximately diagonal in the field amplitude basis in order that one be able to assign definite probabilities to different field amplitudes. Otherwise, off-diagonal elements in the density matrix would imply the existence of coherent superpositions of states with different field amplitudes, in which case no equivalent classical description would apply. The suppression of the off-diagonal elements of the density matrix is called ``decoherence'', and is typically found to occur when one considers the coupling of the fluctuations to an environment (usually provided by additional fields). Thus, given such a coupling to an environment, inflation can successfully describe the origin of all structure in the universe out of initial quantum fluctuations.

However, even though inflation has many compelling features, it also has important open problems -- for a comprehensive recent discussion see \cite{Ijjas:2013vea}.
For this reason it is certainly of interest to explore the properties of alternative cosmological models. A particularly attractive alternative to inflation is provided by cyclic models of the universe, containing an ekpyrotic phase \cite{Khoury:2001wf,Steinhardt:2001st,Lehners:2008vx}. This is a phase of slow contraction of the universe, which also renders the universe spatially flat while producing nearly scale-invariant density perturbations (again via amplification of quantum fluctuations). In order to provide a truly viable alternative to inflation, it is necessary that the perturbations produced in ekpyrotic/cyclic models can be reinterpreted as a statistical distribution of classical perturbations. In this paper, we show that, for the currently best-understood ekpyrotic models, this is indeed so. 

\begin{figure}[t]
\begin{center}
\includegraphics[width=0.75\textwidth]{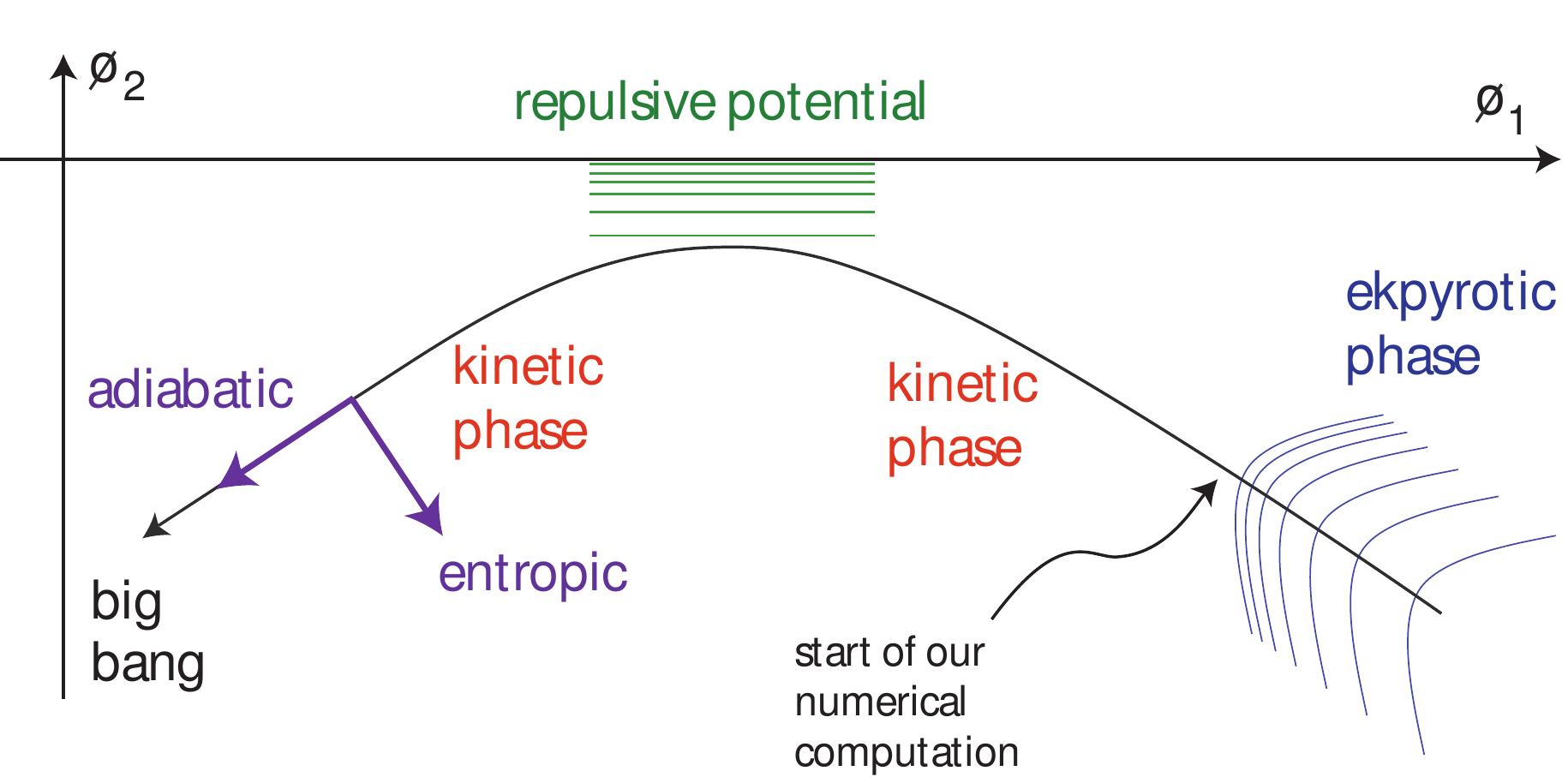}
\caption{\label{Fig1} {\small In the models that we study, the evolution starts out with an ekpyrotic contracting phase to the right of the figure. During this phase, nearly scale-invariant entropy perturbations, which are perturbations transverse to the background trajectory, are amplified and evolve into a highly squeezed quantum state. Such a squeezed quantum state is approximately classical in the sense that it can equivalently be described as a statistical mixture of classical perturbations. After the ekpyrotic phase, the trajectory in scalar field space enters the kinetic phase and bends - this bending causes the conversion of entropy into curvature perturbations, with the latter inheriting both their spectrum and their near-classicality from the entropy perturbations. Moreover, the interactions of entropy and curvature modes during the bending phase cause decoherence to occur, such that the resulting curvature perturbations can be assigned definite classical probabilities for their amplitude. In this way, the ekpyrotic phase produces an ensemble of nearly scale-invariant classical density perturbations in the approach to the bounce.
}}
\end{center}
\end{figure}

These models, which are in good agreement with the observations of the Planck satellite \cite{Lehners:2013cka}, employ the {\it entropic} mechanism for producing density perturbations \cite{Finelli:2002we,Notari:2002yc,Lehners:2007ac}. In these models one considers gravity minimally coupled to two scalar fields with potentials. The perturbations are created via a two-step process. First, during the slowly contracting ekpyrotic phase, nearly scale-invariant entropy perturbations are generated. These entropy perturbations evolve into a highly squeezed quantum state during this generation phase. Subsequently, in the approach to the bounce, the ekpyrotic potential becomes unimportant, and the universe enters a phase in which the energy density is mostly comprised of the kinetic energy of the scalar fields. During this phase, the field space trajectory describing the dynamics undergoes a bend (see Fig. \ref{Fig1}). As a result, the entropy perturbations are converted into adiabatic curvature perturbations which inherit both their squeezed state and their spectrum. What we find in addition is that during this second phase the interactions of entropic and adiabatic modes are already sufficient to cause the density matrix to decohere efficiently, without the need to couple these perturbations to an additional environment. The end result is that the produced curvature perturbations can indeed be faithfully described as a statistical mixture of classical perturbations. Thus, an ekpyrotic phase can equally well provide the seeds for the large-scale structure in our universe, and should be regarded on the same footing as inflationary models in this respect.

The plan of our paper is as follows: in section \ref{section:generation}, we will discuss the generation of adiabatic and entropic fluctuations during an ekpyrotic phase, including a detailed study of the issues of amplification and squeezing. In section \ref{section:conversion}, we will then analyze how the entropy modes become a source for the adiabatic modes, as a result of which the latter ones inherit the classicality properties of the former ones. We also quantify the decoherence of the reduced density matrix that occurs simultaneously. After our discussion section, we include a technical appendix where we present the general formalism for treating two-scalar-field models quantum mechanically, and we highlight potential ambiguities that can arise when performing integrations by parts on the Lagrangian of the system under consideration. A second appendix lists a number of useful formulae involving Bessel functions.

\section{Generation of cosmological perturbations and squeezing} \label{section:generation}

The model that we discuss involves gravity minimally coupled to two scalar fields, with action
\be S=\int
\sqrt{-g} \left[ \frac{R}{2}-\frac{1}{2}(\pt\phi_1)^2
-\frac{1}{2}(\pt\phi_2)^2-V(\phi_1,\phi_2) \right].\ee
We are assuming that during the
ekpyrotic phase, both fields have (steep and negative) ekpyrotic-type potentials,
{\it i.e.} \be V(\phi_1,\phi_2) =-V_1 e^{-c_1 \phi_1} - V_2
e^{-c_2 \phi_2}. \label{potential2field}\ee  Then the analysis is greatly simplified by rotating to the new fields $\s$ and $s$ pointing transverse and perpendicular to
the field velocity respectively
\cite{Koyama:2007mg,Koyama:2007ag}. It is  convenient
to first introduce the  angle $\theta$ in field space, defined by \cite{Gordon:2000hv}
\be
\cos \theta \equiv \frac{\dot{\phi}_1}{\sqrt{\dot\phi_1^2+\dot\phi_2^2}}\, ,
\quad \quad \sin \theta \equiv
\frac{\dot{\phi}_2}{\sqrt{\dot\phi_1^2+\dot\phi_2^2}}\, .
\ee
Then, if we write the fields together as $\phi_J=(\phi_1,\phi_2)$ the adiabatic and entropy directions are defined respectively by the vectors
\be
e_\sigma^J =
\left(\cos \theta,\sin \theta  \right),
\qquad e_s^J =
\left(-\sin \theta , \cos \theta \right).
\ee
In terms of
these new variables, the potential can be re-expressed as \be
V=-V_0\, e^{\sqrt{2\e}\s} \left[1+\k_2 \e s^2+\cdots \right],
\label{potentialParameterized}\ee with $1/\e=2/c_1^2 + 2/c_2^2$ and $V_0$ a constant. For exact exponentials
of the form (\ref{potential2field}), one has
$\k_2=1$, which indicates that if we slightly extend the class of potentials we consider we may take $\k_2$ to be close to $1$ (as we will see below, in such a case the spectral index of the entropy perturbations will be close to scale-invariant \cite{Buchbinder:2007tw,Lehners:2013cka}). The ellipsis denotes higher-order terms in $s$ in the potential, which determine the non-gaussian corrections to the primordial perturbations -- these are discussed in detail in \cite{Lehners:2007wc,Lehners:2008my,Lehners:2009ja}, for a review see \cite{Lehners:2010fy}. The
ekpyrotic scaling solution is given by \be a(t)=(-t)^{1/\e} \qquad
\s=-\sqrt{\frac{2}{\e}}\ln \left(-\sqrt{\frac{\e^2 V_0}{\e-3}} t\right) \qquad
s=0, \label{ScalingSolution}\ee where time runs from large negative to small negative values, and with the angle $\th$ being
constant. The solution corresponds to motion along a ridge in
the potential and thus the two-field ekpyrotic
background evolution is unstable
\cite{Lehners:2007ac,Tolley:2007nq}. This instability has significant consequences: it determines the global structure of a cyclic universe employing the entropic mechanism (this is discussed in detail in \cite{Lehners:2008qe,Lehners:2011ig} -- see also the essay \cite{Lehners:2009eg}), and it is also responsible for amplifying the quantum perturbations in the entropic direction.

In order to discuss the quantum fluctuations, we must expand the action to second order in fluctuations. The gauge-invariant fluctuations that we are interested in are the comoving curvature perturbation $\mathcal{R}$ and the entropy perturbation $\delta s.$ In comoving gauge, the curvature perturbation is defined as a space-time dependent fluctuation in the scale factor
\be
ds^2 = - dt^2 + a(t)^2 e^{-2\mathcal{R}(t,\underline{x})} d \underline{x}^2,
\ee while the entropy perturbation is defined (in any gauge) as
\be
\delta s = e_s^J \delta \phi_J = \cos \theta \, \delta \phi_2 - \sin \theta \, \delta \phi_1.
\ee
In fact, the re-scaled versions
\be
v_\sigma = z \mathcal{R}, \, v_s = a \delta s, \quad \textrm{with} \, \, z=\frac{a \dot\s}{H},
\ee
turn out to be the canonically normalized variables. We will also switch to conformal time $\tau$ (with $d t = a\, d\tau$), and denote derivatives w.r.t. conformal time with primes. Moreover, we find it most convenient to work in momentum space from the start, but since modes with different wavenumbers $k$ are decoupled, we will for the most part suppress such labels. Then, as shown in \cite{Langlois:2008mn}, the Lagrangian for the real/imaginary parts of the Fourier components of the curvature and entropy perturbations reads
\begin{eqnarray} \label{eq:modeLagrangian}
L & = & \frac{1}{2} v _{\sigma} ^{ \prime 2} + \frac{1}{2} v _{s} ^{ \prime 2} -\frac{z'}{z}v_{\sigma}'v_\sigma - \frac{a'}{a}v_s' v_s - 2 \theta' v _{\sigma}' v _{s} - \frac{1}{2} m _{\sigma} ^2 v _{\sigma} ^2 + 2 \theta' \frac{z'}{z} v _{\sigma} v _{s} - \frac{1}{2} m _{s} ^2 v _{s} ^2 \;,\\
m _{\sigma} ^2 & = & k ^2 - \frac{z^{\prime 2}}{z^2} \;,\\
m _{s} ^2 & = & k ^2 - \frac{a^{\prime 2}}{a^2} + a ^2 V_{ss} - \theta ^{ \prime 2} \;,
\end{eqnarray}
where
\begin{eqnarray}
z &\equiv& a \frac{ \sigma'}{\mathcal{H}},\qquad \mathcal{H}\equiv \frac{a'}{a} ,\qquad \sigma' \equiv \left( \phi_1 ^{ \prime 2} + \phi_2 ^{ \prime 2} \right) ^{1/2} \;,\\
V_{ss} & \equiv & \left(V_{, \phi_1 \phi_1} \phi_2 ^{ \prime 2} - 2 V_{, \phi_1 \phi_2} \phi'_1 \phi'_2 + V_{, \phi_2 \phi_2} \phi_1 ^{ \prime 2} \right)/ \sigma ^{ \prime 2} \;.
\end{eqnarray}
Varying the Lagrangian leads to the linearized equations of motion
\begin{eqnarray} \label{eq:pertsigma}
v_ \sigma '' + \mu_ \sigma ^2 v_ \sigma - \frac{2}{z}(z\theta' v_s)'  & = & 0 \;,\\ \label{eq:perts}
v_s'' + \mu_s ^2 v_s + 2z\theta' \left(\frac{v_\sigma}{z}\right)' & = & 0 \;,
\end{eqnarray}
where
\be
\mu_\sigma^2 = k^2 - \frac{z''}{z}, \quad \mu_s^2 = k^2 - \frac{a''}{a} + a^2 V_{ss} - \theta^{\prime 2}.
\ee
The canonical momenta are
\be 
\pi_\sigma = v_{\sigma}' -\frac{z'}{z} v_\sigma -2\theta' v_s,  \quad \pi_s = v_s' -\frac{a'}{a} v_s, \label{Canonicalmomenta}
\ee
and consequently the Hamiltonian is given by
\bea
H & = & \frac{1}{2} \left( \pi _{\sigma} + \frac{z'}{z} v_\sigma +2\theta' v_s \right) ^2 + \frac{1}{2} \left( \pi _{s} +\frac{a'}{a} v_s \right) ^2  \\ \nonumber
&& + \frac{1}{2} m _{\sigma} ^2 v _{\sigma} ^2 + \frac{1}{2} m _{s} ^2 v _{s} ^2 -2\theta'\frac{z'}{z} v _{\sigma} v_{s} \;.
\end{eqnarray}
We can quantize the perturbations as usual by promoting the fields to operators. In our case, the general solution of the resulting Heisenberg equations can be written in terms of two sets of creation/annihilation operators:
\bea
\hat{v}_\sigma &=& f_\sigma \hat{a} + f_{\sigma}^{*}\hat{a} ^{\dagger} + g_\sigma \hat{b} + g_{\sigma}^{*} \hat{b} ^{\dagger} \;, \\
\hat{v}_s &=& f_s \hat{a} + f_s^{*}\hat{a} ^{\dagger} + g_s \hat{b} + g_s^{*} \hat{b} ^{\dagger} \;,
\eea
where $f_{\sigma,s}$ and $g_{\sigma,s}$ are time-dependent, complex, linearly independent solutions of the equations of motion. Analogous expressions for the momentum operators follow from Eq. (\ref{Canonicalmomenta}).
The following quantities (Wronskians) are constants of motion,
\bea
f_\sigma (f_{\sigma}^{*\prime}-\frac{z'}{z} f_{\sigma}^* -2\theta' f_s^*) + f_s (f_s^{*\prime} - \frac{a'}{a} f_s^*) - c.c. &=& i, \label{eq:wronskian1}\\
g_\sigma (g_{\sigma}^{*\prime}-\frac{z'}{z} g_{\sigma}^* -2\theta' g_s^*) + g_s (g_s^{*\prime} - \frac{a'}{a} g_s^*) - c.c. &=& i, \\
f_\sigma (g_{\sigma}^{\prime}-\frac{z'}{z} g_{\sigma} -2\theta' g_s) + f_s (g_s^{\prime} - \frac{a'}{a} g_s) - (f \leftrightarrow g) &=& 0, \\
f_\sigma (g_{\sigma}^{*\prime}-\frac{z'}{z} g_{\sigma}^* -2\theta' g_s^*) + f_s (g_s^{*\prime} - \frac{a'}{a} g_s^*) - (f \leftrightarrow g) &=& 0, \label{eq:wronskian4}
\eea
where we have fixed the right hand sides in such a way as to ensure the canonical normalization of the mode functions.
Using these Wronskian relations, it is then possible to re-express the annihilation operators $\hat{a},\hat{b}$ in terms of the mode functions and their canonical momenta:
\begin{eqnarray}
i\, \hat{a} & = & (f_{\sigma}^{*\prime}-\frac{z'}{z} f_{\sigma}^* -2\theta' f_s^*) \hat{v}_\sigma -f_{\sigma}^* \hat{\pi}_\sigma +(f_s^{*\prime} - \frac{a'}{a} f_s^*) \hat{v}_s - f_s^* \hat{\pi}_s \;,\\
i\,\hat{b} & = & (g_{\sigma}^{*\prime}-\frac{z'}{z} g_{\sigma}^* -2\theta' g_s^*) \hat{v}_\sigma -g_{\sigma}^* \hat{\pi}_\sigma +(g_s^{*\prime} - \frac{a'}{a} g_s^*) \hat{v}_s - g_s^* \hat{\pi}_s \;.
\end{eqnarray}
We define the vacuum state as usual by requiring that it vanishes when acted upon by the annihilation operators
\be
\hat{a} \ket{0} = \hat{b} \ket{0} = 0 \;.
\ee
Using the expressions above and the canonical replacements $\pi_{\sigma,s} \rightarrow -i \frac{\pt}{\pt v_{\sigma,s}}$ we can then obtain an expression for the corresponding (Schr\"{o}dinger picture) wavefunction $\Psi:$
\begin{eqnarray}
\Psi(v_\sigma,v_s) & = & N\, \textrm{exp} \left( - \frac{1}{2}A_{ \sigma \sigma} v_ \sigma ^2 - A_{ \sigma s} v_ \sigma v_ s - \frac{1}{2} A_{ s s} v_ s ^2 \right) \;, \label{eq:wavefunction}
\end{eqnarray}
where $N$ is a normalization factor and where the correlators are given by
\begin{eqnarray} \label{eq:solCorr1or}
A_{ \sigma \sigma} & = & - i \frac{ g_{s} ^{*} f_{\sigma}^{*\prime}-f_s^* g_{\sigma}^{*\prime}}{ g_{s}^{*} f_{\sigma}^{*}  - f_{s} ^{*} g_{\sigma}^{*}} + i \frac{z'}{z} \;,\\ \label{eq:solCorr2}
A_{ ss} & = & - i \frac{  f_{\sigma}^{*}g_{s}^{*\prime}  - g_{\sigma}^{*} f_{s}^{*\prime} }{ f_{\sigma}^{*} g_{s}^{*} - g_{\sigma}^{*} f_{s} ^{*} } + i \frac{a'}{a} \;, \\ \label{eq:solCorr3}
A_{ \sigma s} & = & -i \frac{  f_{\sigma}^{*}g_{\sigma}^{*\prime}  - g_{\sigma}^{*} f_{\sigma}^{*\prime} }{ f_{\sigma}^{*} g_{s}^{*} - g_{\sigma}^{*} f_{s} ^{*} } + 2i\theta' = - i \frac{ g_{s} ^{*} f_{s}^{*\prime}-f_s^* g_{s}^{*\prime}}{ g_{s}^{*} f_{\sigma}^{*}  - f_{s} ^{*} g_{\sigma}^{*}} \;.
\end{eqnarray}
The correlators satisfy their own equations of motion, which can be derived either via the equations of motion of the mode functions, or via the time-dependent  Schr\"{o}dinger equation $i \Psi' = \hat{H} \Psi$. Both methods lead to
\begin{eqnarray}
i A_{\sigma\sigma}' &=& (A_{\sigma\sigma}-i\frac{z'}{z})^2 + A_{\sigma s}^2 - m_{\sigma}^2 \;,\\
i A_{ss}' &=& (A_{ss}-i\frac{a'}{a})^2 + (A_{\sigma s}-2i\theta')^2 - m_{s}^2 \;, \\
i A_{\sigma s}' &=& A_{\sigma s}(A_{\sigma\sigma} + A_{ss}) - i(\frac{z'}{z}+\frac{a'}{a})A_{\sigma s} -2i\theta' A_{\sigma \sigma} \;.
\end{eqnarray}

Having set up the necessary formalism, we can now apply it to the ekpyrotic phase, during which the following relations hold:
\begin{eqnarray}
\sigma & = & - \sqrt{ \frac{2}{ \epsilon}} \ln{ \left( - \sqrt{ \frac{ \epsilon ^2 V_0}{ \epsilon - 3}}t \right)}, \quad
s  =  0 \;, \\
a & = & a_0 (- t) ^{1/ \epsilon} = \bar{a}_0(- \tau) ^{1/( \epsilon-1)} \;,\\
  \frac{a''}{a} & = & \frac{z''}{z} = - \frac{ \epsilon -2}{( \epsilon -1) ^2} \frac{1}{ \tau ^2} \;, \\
  \frac{a''}{a} - a ^2V_{ ,ss} & \approx & (2\k_2 - \frac{2\k_2}{\e} - \frac{1}{\e}) \frac{1}{ \tau ^2}
   \;.
\end{eqnarray}
Hence the mode equations \eqref{eq:pertsigma} and \eqref{eq:perts} read
  \begin{eqnarray}
  v'' + \left(k ^2 + \frac{\frac{1}{4}-\alpha^2}{ \eta ^2} \right) v & = & 0 \label{eq:modefct}\;,
  \end{eqnarray}
  with
  \be \label{eq:alpha}
  \a_ \sigma^2  =  \frac{ (\epsilon - 3)^2}{4( \epsilon - 1) ^2} \;, \quad
  \a _s^2  \approx  \frac{1}{4} + 2\k_2 - \frac{2\k_2}{\e} - \frac{1}{\e} \;.
  \ee
Note that during the ekpyrotic phase $\theta'=0$ and thus there is no coupling between adiabatic and entropic modes. Consequently, the two modes can be treated independently. The solutions respecting the Wronskian conditions may be written in terms of Bessel functions,
\be
v  =  \sqrt{ \frac{ \pi}{4k}} \sqrt{- k \tau} \Big( J_{ \alpha}(- k \tau) + i\, Y_{ \alpha}(- k \tau) \Big) \;.
\ee
When $ \epsilon$ is large and $|\k_2 -1| \ll 1$ one has
\begin{eqnarray} \label{eq:alphaSigma}
\alpha_ \sigma & \approx &  \frac{1}{2} - \frac{ 1}{ \epsilon} \;,\\ \label{eq:alphas}
\alpha_ s & \approx & \frac{3}{2} + \frac{2}{3}(\k_2 -1) - \frac{1}{ \epsilon}  \;,
\end{eqnarray}
where these solutions apply to the two non-zero mode functions $f_\s$ and $g_s,$ while $f_s$ and $g_\s$ are zero during the ekpyrotic phase. Using the asymptotic behaviors of Bessel functions provided in Appendix \ref{section:Bessel}
we obtain
\begin{eqnarray}
f_{\sigma} & \simeq & \frac{ \Gamma( \alpha) 2 ^{ \alpha}}{ \sqrt{ 4 \pi k}} \left( \frac{ \pi}{ 2 ^{2 \alpha} \Gamma( \alpha) \Gamma( \alpha + 1)} (-k\tau) ^{1 - 1/ \epsilon} - i\, (-k\tau) ^{1/ \epsilon} \right) \label{eq:modefctfsigma}\;,\\
g_s & \simeq & \frac{ \Gamma( \alpha) 2 ^{\alpha}}{ \sqrt{ 4 \pi k}} \left( \frac{ \pi}{ 2 ^{2 \alpha} \Gamma( \alpha) \Gamma( \alpha+1)} (-k\tau)^{2 +2(\kappa_2 -1)/3- 1/ \epsilon} - \frac{i}{(-k\tau)^{1 + 2(\kappa_2 - 1)/3-1/ \epsilon}} \right)\label{eq:modefctgs} \;.
\end{eqnarray}
We are now in a position to evaluate the correlators during the ekpyrotic phase. Again using the formulae of Appendix \ref{section:Bessel} as well as the relation $(1- 2 \alpha_ \sigma)( \epsilon - 1)=2$, we find that the correlators are given by
\bea
A_{\s\s} &\approx& \frac{k ^{2 \alpha_\sigma}}{ 2 ^{ 2 \alpha_\s-1} \Gamma( \alpha_\s)} | \tau| ^{2 \alpha_\s - 1} \left( \frac{\pi}{\Gamma(\a_\s)} - i \a_\s \Gamma( - \alpha_\s) \cos( \pi \alpha_\s) \right), \label{eq:corr_ek_1} \\
A_{\s s} &=& 0 \\
A_{ss} &\approx&  \frac{ \pi k ^{2 \alpha_s}}{ 2 ^{ 2 \alpha_s - 1}\Gamma( \alpha_s) ^2} | \tau| ^{ 2 \alpha_s - 1} - i  \frac{\left(\alpha_s - \frac{1}{2}+ \frac{1}{\epsilon} \right)}{|\tau|}  \;,
\eea
where $\alpha_{\sigma}$ and $\alpha_s$ were listed above in Eqs. (\ref{eq:alphaSigma})-(\ref{eq:alphas}), and where we have kept only the leading real and imaginary parts. One can immediately see that the adiabatic and entropic correlators behave very differently: let us first look at the adiabatic modes, which are characterized by a blue spectrum $n_s = 4-2\a_\sigma \approx 3.$ The criterion for describing a quantum state as being semi-classical (in a WKB sense) is that the phase of the wavefunction must
vary much faster than its amplitude. From \eqref{eq:wavefunction} we can see that this corresponds to the criterion that the imaginary part of the correlator must be much larger than the real part. However, in the present case, the real and imaginary parts of $A_{\s\s}$ have the same time dependence, and thus their relative magnitude remains fixed over time. Moreover, as the explicit expression \eqref{eq:corr_ek_1} shows, their magnitudes are of the same order
\be
\textrm{Re}(A_{\s\s}) \approx \textrm{Im}(A_{\s\s})
\ee
at all times, and hence these blue modes cannot be given a classical interpretation. This calculation reproduces the results of \cite{Tseng:2012qd}. We note that the correlator becomes large as $\tau \rightarrow 0^-$, which implies that the dispersion of the $v_\s$ modes becomes small as the ekpyrotic phase progresses. Thus there occurs no significant production of these modes.

By contrast, the real part of the entropic correlator $A_{ss}$ becomes small as $\tau \rightarrow 0^-.$ Hence in this case the dispersion of the entropic perturbations becomes large -- in other words, such modes are amplified as the ekpyrotic phase proceeds. Moreover, the imaginary part of the correlator becomes large in magnitude, so that the phase of the wavefunction evolves much faster than its amplitude \footnote{It is interesting to note that the leading imaginary term of the correlator $A_{ss}$ does not appear in $A_{\s\s},$ where it cancels out exactly due to Eq. (\ref{eq:alpha}).}. Over time such modes behave increasingly classically in a WKB sense, with
\be
\frac{|\textrm{Im}(A_{ss})|}{| \textrm{Re}(A_{ss})|} \approx \frac{1}{|k \tau|^{2 \alpha_s}} \gg 1 \, \textrm{as} \, |k\tau| \ll 1.
\ee
This formula shows that the perturbation modes evolve into a highly squeezed Gaussian state as they leave the horizon, in complete analogy with inflationary perturbations. The entropic modes of interest to us belong to this category (with $\a_s \approx \frac{3}{2}).$ Their spectral index  is given by
\be
n_s = 4-2\a_s \approx 1 - \frac{4}{3}(\k_2 -1) + \frac{2}{\e},
\ee
and thus their spectrum is nearly scale-invariant spectrum when $\epsilon \gg 1$ and $|\kappa_2-1| \ll 1$. All of these features, with one important exception, make these modes suitable candidates for producing the seeds of the large-scale structure in the universe. The exception is of course that these modes correspond to local perturbations in the entropy, whereas observations indicate that the temperature fluctuations in the cosmic background radiation are primarily due to adiabatic curvature fluctuations. As we will see in the next section, it is precisely the process that converts such entropic into adiabatic fluctuations that is also responsible for decohering them, thus confirming their classical appearance.

\section{Conversion of entropic into adiabatic perturbations and decoherence} \label{section:conversion}

In the previous section, we have shown that the ekpyrotic phase produces two sets of fluctuations: adiabatic perturbations with a blue spectrum, a small amplitude and no classical interpretation, alongside nearly scale-invariant entropic modes in a highly squeezed semi-classical state. In the entropic mechanism, after the ekpyrotic phase has come to an end the entropy modes get converted into adiabatic curvature perturbations. This conversion occurs when the trajectory in field space undergoes a bend, as shown in Fig. \ref{Fig1}. (As described in \cite{Lehners:2006pu}, such a bending of the trajectory occurs automatically in the embedding of ekpyrotic models into heterotic M-theory.) In the following, we would like to examine to what extent the semi-classical properties of the entropic modes get inherited by the adiabatic modes during this process. In the model that we are studying, after the conversion process the universe briefly remains in a phase dominated by the kinetic energy of the scalar fields before it undergoes a bounce into an expanding hot big bang phase (without inflation occurring). The idea is that the bounce is not completely elastic, and that a small fraction of the energy density gets converted into radiation and matter degrees of freedom during the bounce phase \cite{Steinhardt:2001st}. (Models describing either classically singular or non-singular bounces form an active research topic, see {\it e.g.} \cite{Turok:2004gb,Buchbinder:2007ad,Creminelli:2007aq,Lehners:2011kr,Cai:2012va,Osipov:2013ssa,Qiu:2013eoa}.) The curvature perturbation corresponds to a local fluctuation in the scale factor of the universe, and thus it has the effect that the bounce will occur at slightly different times in different regions of the universe. In this way, any radiation and matter that get produced during the bounce inherit their perturbations directly from the curvature perturbation.

\subsection{The reduced density matrix and decoherence}
We are assuming here that the entropy field decays sometime after the ekpyrotic phase has come to an end, e.g. during reheating at the bounce. Then the only observable imprint from the ekpyrotic phase on the cosmic background radiation is represented by the adiabatic curvature perturbations - concerning this point see also the analogous discussion in the context of two-field inflation in \cite{Prokopec:2006fc}. This means that we should treat the entropic modes as an inaccessible environment for the curvature modes, and hence, in order to characterize the generated curvature perturbations, we must study the reduced density matrix that one obtains by tracing over the entropic degrees of freedom. As we will see, this reduced density matrix allows us to quantify the effective classicality of the curvature perturbations. Explicitly, it is given by
\begin{eqnarray}
\rho( v_ \sigma,\bar{v}_ \sigma ) &=& \langle v_\sigma | \textrm{Tr}_{v_s} \hat{\rho} | \bar{v}_\sigma \rangle \\ & = & \int dv_s \Psi(v_\sigma,v_s) \Psi^*(\bar{v}_\sigma,v_s) \\
&=& \tilde{N} \,\textrm{exp} \left( - \frac{1}{2} C_{SS} v_S ^2 - \frac{1}{2} C_{DD} v_D ^2 - i \, C_{SD} v_S v_D \right) \;,
\eea
where $\tilde{N}$ is a normalization factor and $\hat{\rho}$ denotes the full density matrix. For simplicity, we will adhere to common practice and call the left-hand side $\rho$ the reduced density matrix, although it actually only corresponds to one particular element thereof. Here we have defined
\bea
v_S & \equiv & \frac{1}{2} \left(v_ \sigma + \bar{v} _{\sigma} \right) \;,\\
v_D & \equiv & v_ \sigma - \bar{v} _{\sigma} \;,
\eea
while the correlators are given by \cite{Prokopec:2006fc}\footnote{In \cite{Prokopec:2006fc} an additional erroneous factor of 2 is present inside the brackets in the expression of $C_{SD}$. Our preliminary analysis shows that the use of the correct expression \eqref{eq:exprCSD} does not significantly change the conclusions of that work, while it plays a crucial role in the results of the present paper.}
\bea
C_{SS} & = & 2 A _{ \sigma \sigma} ^{R} \left( 1 - \frac{( A _{ \sigma s} ^{R})^2}{A_{ss} ^{R} A _{ \sigma \sigma} ^{R}} \right) \;,\\ \label{eq:exprCSD}
C_{SD} & = & A_{ \sigma \sigma} ^{I} \left(1 - \frac{A _{ \sigma s} ^{I} A_{ \sigma s} ^{R}}{A_{ s s} ^{R}A_{ \sigma \sigma} ^{I}} \right)\;,\\
C_{DD} & = & \frac{1}{2} A_{ \sigma \sigma} ^{R} \left( 1 + \frac{( A_ { \sigma s} ^{I}) ^2}{ A_{ss} ^{R} A_{ \sigma \sigma} ^{R}} \right) \;.
\end{eqnarray}
Note the extra factor of $i$ that we have pulled out of the coefficient of the mixed $v_S v_D$ term, such that $C_{SD}$ is real. 

The situation that we are aiming for is one where the density matrix is approximately diagonal in the field amplitude basis - in this case we say that the density matrix has {\it decohered}. More explicitly, we would like the density matrix to yield a sizeable probability when we choose the field amplitudes $v_\sigma$ and $\bar{v}_\sigma$ to be equal, but a zero or very small probability when they are unequal, $v_\sigma \neq \bar{v}_\sigma.$ In this case, the density matrix describes with high accuracy a statistical mixture of states with definite field amplitudes, and thus the quantum perturbations that we are studying can then equivalently be described as an ensemble of classical density perturbations\footnote{If the density matrix is not diagonal in the field amplitude basis, then the fluctuation modes are in coherent superpositions of states with different field amplitudes. Evidently, such a situation cannot be described classically.}. We can specify the amount of decoherence by evaluating the so-called {\it entanglement entropy}
\be
s_k = \frac{1}{2} \ln \left( \frac{4 C_{DD}}{C_{SS}} \right),
\ee
which quantifies the extent to which the difference terms $v_D$ are suppressed relative to the $v_S$ terms. The entanglement entropy $s_k$ does not obey a simple evolution equation, and one must in fact evaluate the correlators $C_{SS},C_{DD}$ directly in terms of the original correlators $A_{\s\s},A_{\s s}$ and $A_{ss},$ and these in turn are most easily evaluated via their dependence on the mode functions $f_{\s,s}, g_{\s,s}.$ Although approximation techniques exist in order to solve for the evolution of the mode functions during the conversion phase \cite{Lehners:2009qu}, we have found these to be insufficiently accurate for our present purposes, and thus we have solved for the mode functions numerically. In analogy to the numerical calculations performed in \cite{Lehners:2009ja} in the context of non-gaussian corrections to the perturbations that we are studying here, we model the bending of the trajectory during the conversion phase by assuming a repulsive potential
\be
V_{rep} = \frac{\tilde{V}}{\phi_2^2}e^{-(10\phi_2)^2} + V_0,
\ee
where $\tilde{V},V_0$ are small constants (we chose the explicit values $\tilde{V}=8 \cdot10^{-9}, V_0=10^{-12}$). Thus the trajectory bends in the vicinity of the $\phi_2=0$ line. We have added the small constant term $V_0$ in order to improve the numerical stability of our computations. Compared to \cite{Lehners:2009ja}, we have also added an exponential suppression term, which ensures that away from the $\phi_2=0$ line the potential quickly reaches a constant value. This implies that at the start of our computation, the angle of the trajectory in scalar field space is almost exactly constant (with our initial conditions, the starting value of the rate of change of the angle is $\theta'\approx 10^{-36}$). This allows us to see the onset of decoherence with high precision. We have started our numerical evaluation right after the ekpyrotic phase has come to an end, when the background dynamics becomes dominated by the kinetic energy of the scalar fields (incidentally, even during the bending the energy density still remains dominated by the kinetic energy of the scalars). As initial conditions for the mode functions $f_\sigma$ and $g_s$ we have used the explicit analytic expressions (\ref{eq:modefctfsigma}) and (\ref{eq:modefctgs}). During the ekpyrotic phase, the other two mode functions, $f_s$ and $g_\sigma,$ are zero. This is consistent as long as $\theta'=0,$ but since $\theta'$ assumes a (tiny) non-zero value already at the start of our computation, we have also set
\be
f_s'(\tau_0) = -2 \theta' f_\sigma (\tau_0), \quad g_\sigma(\tau_0) = 0
\ee
such that the Wronskians (\ref{eq:wronskian1})-(\ref{eq:wronskian4}) remain exactly satisfied at the starting time $\tau_0$.

\begin{figure}[t]
\begin{minipage}{\smallWidthLeft}
\flushleft
\includegraphics[width=\smallWidthRight]{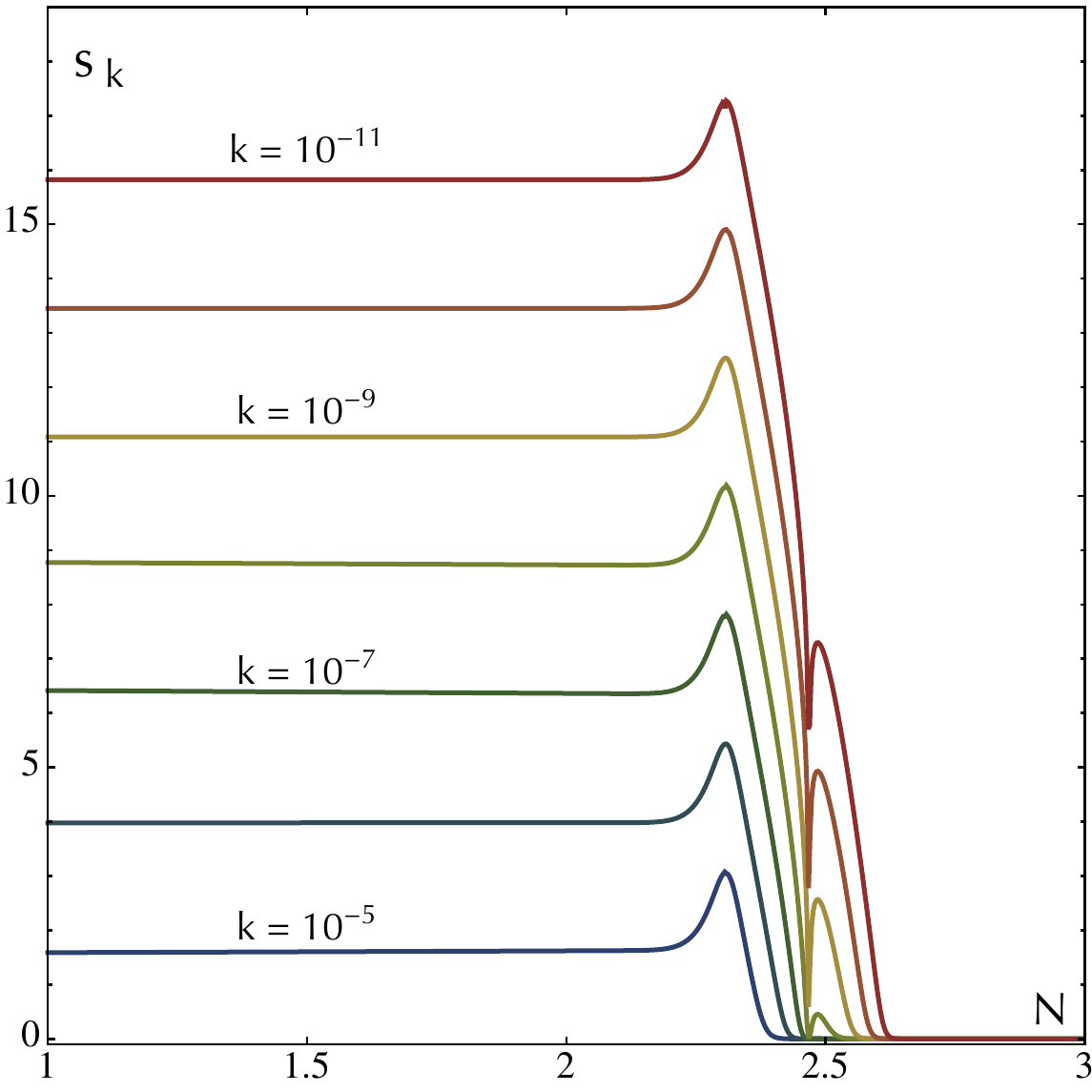}
\end{minipage}%
\begin{minipage}{\smallWidthRight}
\flushleft
\includegraphics[width=\smallWidthRight]{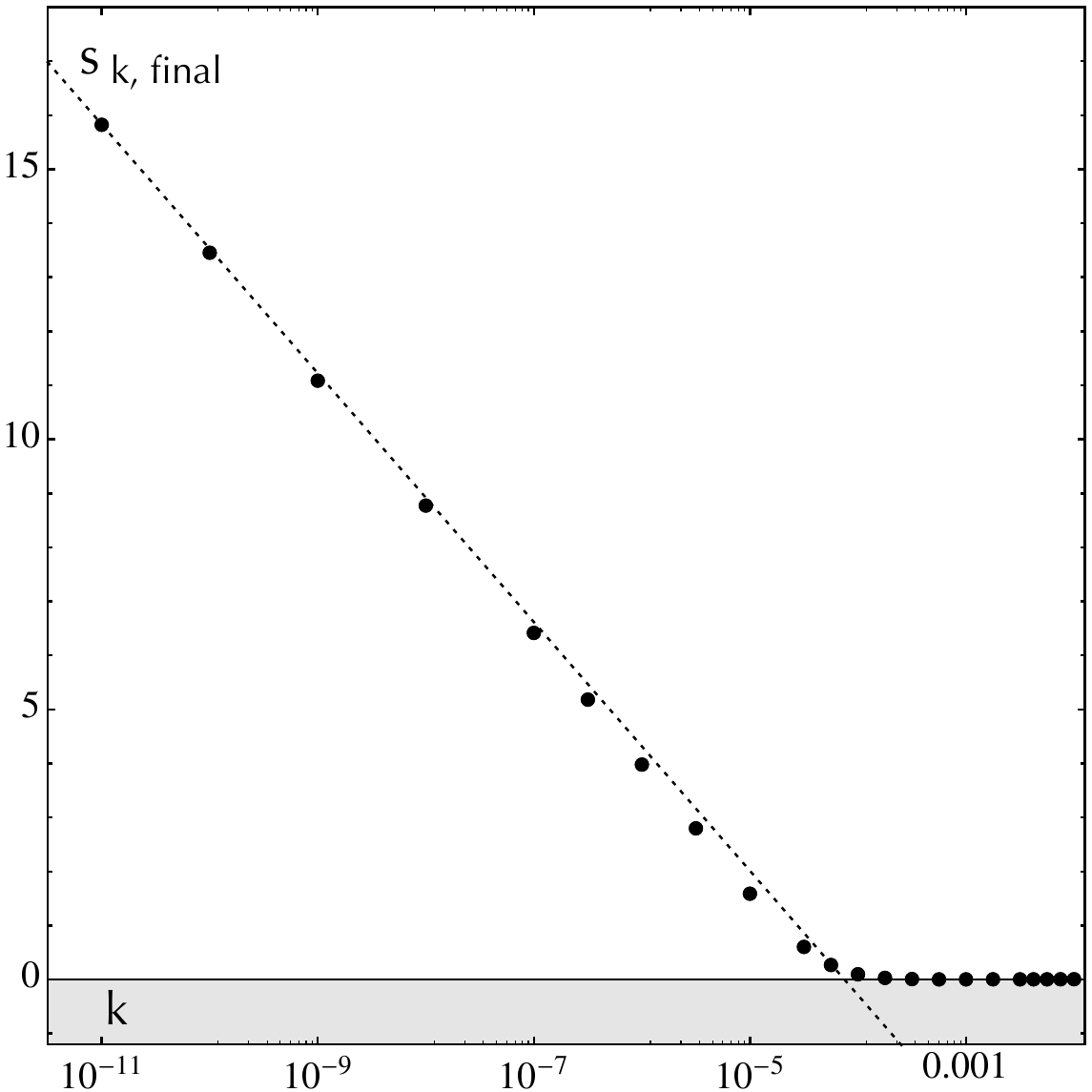}
\end{minipage}
\caption{\small \label{FigEntropy} These figures show the entanglement entropy $s_k$ as a function of time for various wavenumbers $k.$ Time runs from the right to the left, and is given in units of scale-factor time $N \propto \ln(a).$ Left panel: the perfect quantum coherence at the end of the ekpyrotic phase is rapidly destroyed during the conversion process, and reaches a constant value in the approach to the bounce. Right panel: for small--wavelength modes, the final amount of decoherence $s_{k,final}$ can be seen to be inversely proportional to $k$ -- the dashed line corresponds to the fitting formula (\eqref{entropyfitting}).}
\end{figure}

Fig. \ref{FigEntropy} then shows our results for the evolution of the entanglement entropy $s_k$ as a function of scale-factor time $N \propto \ln a.$ Since the universe is contracting, $N$ is decreasing, and hence in the figure time is running from right to left. The figure shows how, for different values of the wavenumber $k$, we start with perfect quantum coherence on the right, and then, as the conversion process takes place the environment provided by the entropy modes effectively decoheres the reduced density matrix. As is evident from the figure, the entanglement entropy reaches a constant value after the conversion process has ended. As mentioned above, the entanglement entropy $s_k$ does not obey a simple evolution equation, as its evolution is governed by
\be s_k'=-2\theta'\frac{A_{\s\s}^R A_{\s s}^R + A_{\s\s}^I A_{\s s}^I}{A_{\s\s}^R A_{ss}^R + (A_{\s s}^I)^2}, \ee
but this explicit formula immediately confirms that $s_k$ is constant when $\theta'=0.$ From a physical point of view, this is also easy to understand, as the adiabatic and entropy modes become decoupled when $\theta'=0.$

An interesting feature of the model we are studying is that the amount of decoherence increases rapidly for longer (comoving) wavelengths $1/k.$ As illustrated in Fig. \ref{FigEntropy}, right panel, the final values of $s_k$ are well fitted by the relation
\be
s_{k,final} \sim -\ln \left( \frac{k}{k_0} \right),\quad k_0 \sim 10^{-4}, \qquad (k \lesssim k_0). \label{entropyfitting}
\ee
The modes of cosmological interest, {\it i.e.} those that we can observe in the cosmic background radiation, have wavenumbers on the order of $k\sim 10^{-25}.$ We have not been able to reach such small values numerically, but if we just extrapolate the fitting formula above, we can estimate that for such modes the entanglement entropy is on the order of $50.$ Recalling that $s_k$ is essentially half of the logarithm of the suppression factor in the reduced density matrix, we see that decoherence is extremely effective for all modes of cosmological interest (thus, for such modes, off-diagonal elements are suppressed by a factor $10^{42}$ relative to the diagonal elements), and that in the present model one does not even have to consider a coupling to an additional environment in order to obtain sufficient decoherence -- the interactions between adiabatic and entropic modes during the conversion phase are entirely sufficient!

\subsection{Amplification, squeezing and semi-classicality}
As we have just seen, the reduced density matrix decoheres very effectively during the conversion phase. Hence, at the end of that process, we are left with adiabatic curvature perturbations that have a nearly scale-invariant spectrum and that behave like an ensemble of classical perturbations with definite amplitudes. However, it would still be interesting to know precisely to what extent the curvature perturbations have become classical: to what extent has the squeezed state of the entropic perturbations been inherited by the curvature perturbations? In other words, to what extent is the field momentum correlated with the amplitude according to the classical relation? This question is relevant for cosmological applications, as the acoustic peaks in the cosmic background radiation show that the momentum of the perturbations must have been highly correlated with the field amplitudes. 

We can address this question by looking at the Wigner function. As is well-known, due to the uncertainty relations, it is impossible in quantum mechanics to talk about a precisely defined phase space. However, for semi-classical states an effective phase space description becomes available by making use of quasi-probability distributions, of which the Wigner function is the best-known example - for a review see \cite{Hillery:1983ms}. The Wigner function for $v_\sigma,\pi_\sigma$ can be obtained from the reduced density matrix via
\bea
W(v,\pi) &=& \frac{1}{2\pi} \int dv_D\, \rho(v-\frac{v_D}{2},v+\frac{v_D}{2}) \, e^{i v_D \pi} \\
&=& \frac{C_{SS} ^{1/2}}{2 \pi C_{DD} ^{1/2}}\; \textrm{exp} \left( - \frac{C_{SS}}{2} v ^2 - \frac{1}{2 C_{DD}}(\pi+C_{SD} v)^2  \right),
\eea
where we have imposed that the total probability is normalised to one, $\int dv d\pi W(v,\pi)=1,$ and where we are now dropping the $\sigma$ subscripts when a possible confusion seems unlikely.
We may rewrite the Wigner function as
\be
W \propto \textrm{exp} \left( -\frac{ v ^2 }{2(\Delta v)^2} -\frac{1}{2(\Delta \pi_{cl})^2}  \left(\pi-\pi_{cl}(v) \right)^2 \right).
\ee
This form lets us identify the classical correlation between $v$ and $ \pi$ that describes the quantum state in the optimal way\footnote{By this we mean that the expectation value $\langle (\hat\pi-\lambda \hat{v})^2 \rangle$ is minimized for $\lambda=-C_{SD}.$}
\bea
\pi_{cl}(v) &=& - C_{SD}\, v.
\eea
The dispersions of $ v$ and $ \pi - \pi_{cl}$ can be read off directly from the Wigner function:
\begin{eqnarray}
\Delta v ^2 & \equiv & \langle \hat{ v}  ^2 \rangle = C_{SS} ^{-1} \;,\\
\Delta \pi_{cl} ^2 & \equiv & \langle \left( \hat{ \pi} + C_{SD} \hat{v} \right) ^2 \rangle  = C_{DD} \;.
\end{eqnarray}
Given that $\langle \hat\pi \rangle = 0,$ the total dispersion of momentum is given by
\begin{equation}
(\Delta \pi) ^2 \equiv \langle  \hat{ \pi} ^2 \rangle = \int dv d\pi\, W(v,\pi) \pi^2= \Delta \pi_{cl} ^2 + C_{SD} ^2 \Delta v ^2 = C_{DD} + \frac{C_{SD} ^2}{C_{SS}}.
\end{equation}
In passing, we note that the entanglement entropy is also directly related to the dispersions in field amplitude and momentum via
\begin{equation} \label{eq:areaEllipse}
\frac{1}{2} e^{s_k} = \sqrt{ \frac{ C_{DD}}{C_{SS}}} = \Delta v\, \Delta \pi_{cl}  \geq \frac{1}{2}. 
\end{equation}
Expressed in this form, one can see that the exponential of the entanglement entropy is given by the area of the Wigner ellipse in phase space (see also Figure \ref{FigSqueezing}).

\begin{figure}[t]
\begin{minipage}{\smallWidthLeft}
\flushleft
\includegraphics[width=\smallWidthRight]{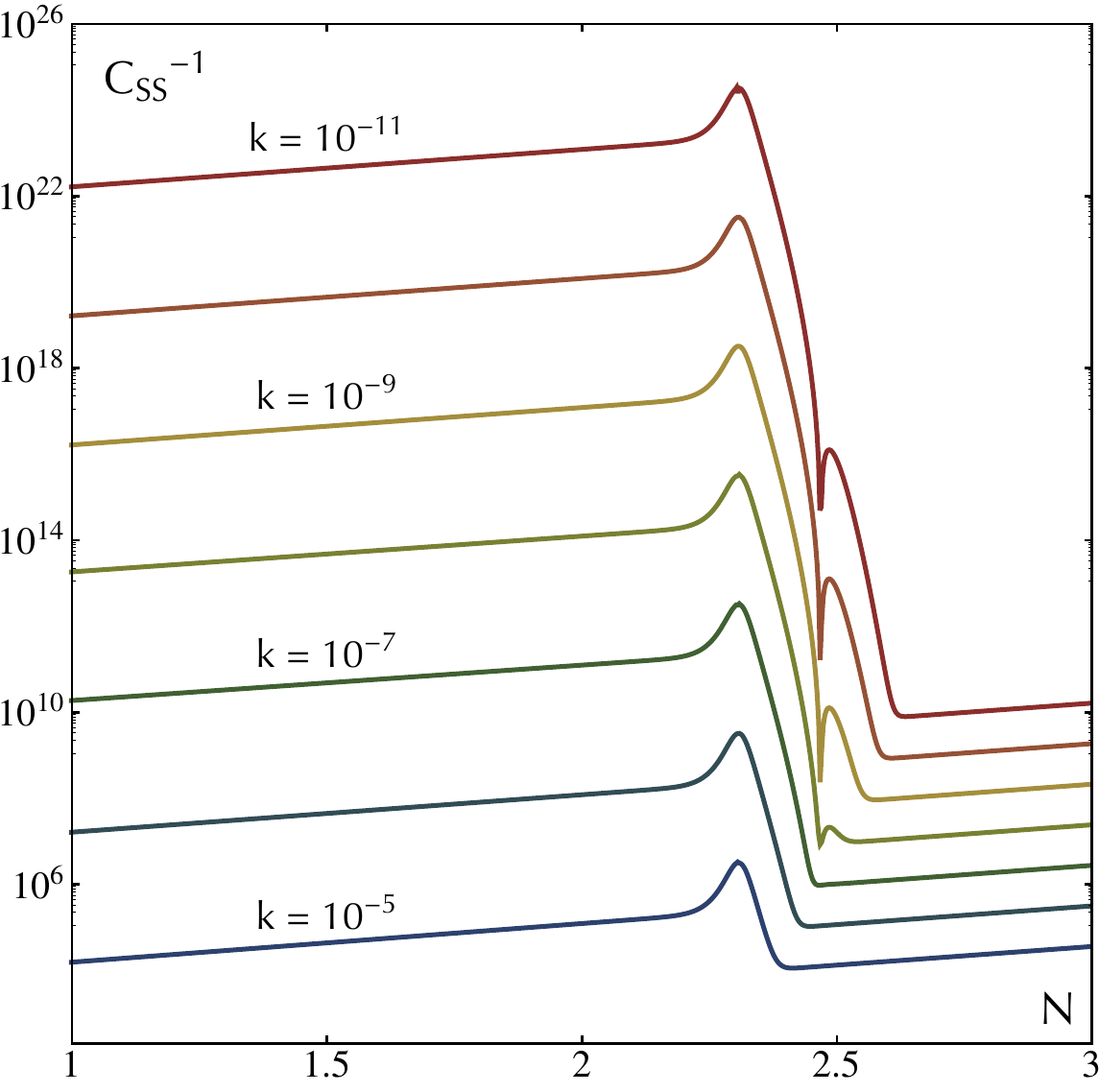}
\end{minipage}%
\begin{minipage}{\smallWidthRight}
\flushleft
\includegraphics[width=\smallWidthRight]{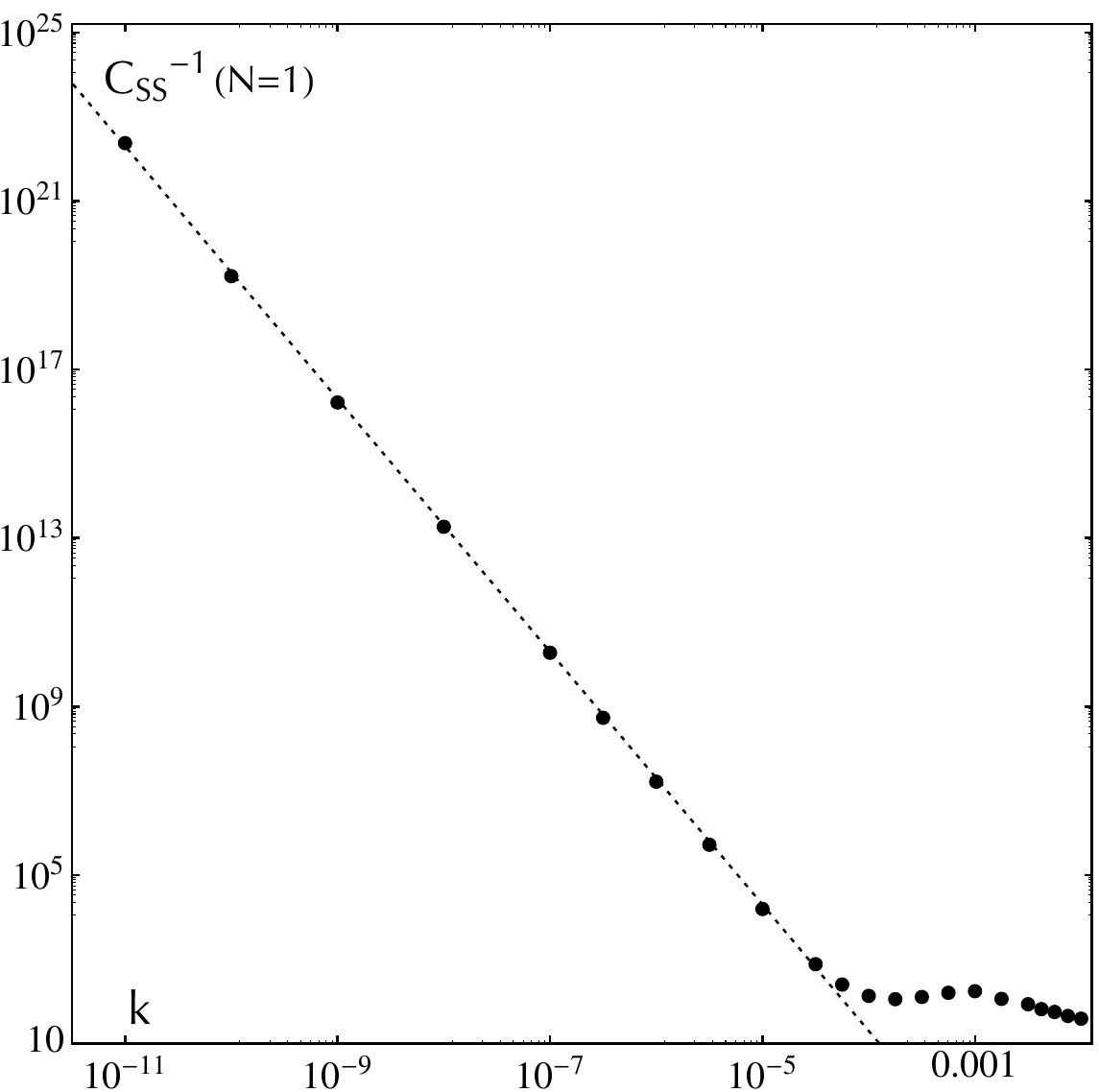}
\end{minipage}
\caption{\small \label{FigExcitation} The figure shows the amplification of the curvature perturbations $(\Delta v)^2 = 1/C_{SS}$
as a function of time (left panel) and at the reference time $N=1$ (right panel) for various wavenumbers $k$. Noting the logarithmic scale on the vertical axis, it is evident that long-wavelength modes become highly amplified during the conversion process. In particular, at fixed time one finds $ ( \Delta v) ^2 \propto k ^{-3}$. During the kinetic phase, this amplification is reduced somewhat, but will remain large if the bounce occurs within not too large a number of e-folds after conversion.
}
\end{figure}

If the dispersion $\Delta v$ is large, this indicates that the curvature perturbations are amplified. In Fig. \ref{FigExcitation} we show our numerical results for the amplification (we have plotted $(\Delta v)^2$) again as a function of the wavenumber $k$. As the figure shows, for small wavenumbers the perturbations are highly amplified during the conversion process, while afterwards, in the approach to the bounce, this amplification is reduced somewhat. If the bounce occurs within a few e-folds after the end of the conversion process, as is natural in the entropic mechanism, then a large amount of amplification remains. The right panel of Fig. \ref{FigExcitation} shows the amplification (plotted here at the reference time $N=1$) as a function of wavenumber. The curve is well fitted by assuming a $k$-dependence 
\be
\Delta v \propto k^{-3/2}, \label{amplifitting}
\ee
which confirms that small-wavelength modes are highly amplified, and that they inherit their spectrum from the entropy modes\footnote{Some reader may not be accustomed to seeing the variance of the perturbations expressed in this way. To provide a link to the usual calculation, note that for a single field the variance is given by $(\Delta v)^2 = 1/C_{SS} = 1/Re(-i v^{*\prime}/v^*) = 1/[-i(v^{*\prime}/v^* - v'/v)] = vv^*,$ where in the last step we have used the wronskian (\ref{eq:wronskian1}).}.

Finally, as also discussed in \cite{Prokopec:2006fc}, it is important to evaluate the degree of classicality of the final adiabatic modes. The effective classicality is largely determined by the dispersion $ \Delta \pi _{cl}$, i.e. by the length of the (typically) shorter axis of the Wigner ellipse (see Figure \ref{FigSqueezing}, top left panel). For this quantity to be small, the dispersion of $ v$ must necessarily be large according to \eqref{eq:areaEllipse}. This implies a large ratio between the lengths of the principal axes of the Wigner ellipse, and is for this reason referred to as \textit{squeezing}. A large amount of squeezing is required in order for the correlations between $v$ and $\pi$ to closely follow their classical counterpart. Combining our definitions and results \eqref{entropyfitting}, \eqref{eq:areaEllipse} and \eqref{amplifitting}, it follows that
\be
\frac{ \Delta \pi _{cl}}{ \Delta v} \leq \frac{\Delta \pi}{\Delta v} \propto k^2
\ee 
and thus on large scales the curvature perturbations are in a highly squeezed state. The crucial question here is whether this effective classicality is such that the produced curvature perturbations lead to the observed features of the cosmic background radiation, and in particular whether they will lead to the observed pattern of peaks and troughs in the angular power spectrum of the cosmic microwave background. These peaks and troughs are caused by acoustic oscillations after the curvature perturbations re-enter the horizon in the expanding phase of the universe. At that point, the evolution of the perturbations is given by circular classical trajectories in the $( k ^{1/2} v, k ^{-1/2} \pi)$ plane. When the Wigner ellipse is highly squeezed in these reduced variables the initial phase (\textit{temporal phase}) of the oscillations is the same for all the modes with the same wavenumber. The classical value of the temporal phase is given by
\begin{equation}
\tan{ \varphi } \equiv  \frac{ |\pi _{cl}(v)|}{ k\,v} = \frac{C_{SD}}{k} \;.
\end{equation}
Moreover, as long as $ \varphi$ is small (which will be the relevant case here), the variance of the temporal phase of a squeezed wavepacket is approximately given by
\begin{equation} \label{eq:deltaPhik}
\Delta \varphi \simeq \frac{ \Delta \pi _{cl}}{ k\, \Delta v } \;.
\end{equation}
Fig. \ref{FigSqueezing} presents numerical results for this indicator. As shown in the bottom left panel, the variance of the temporal phase decreases dramatically during the conversion phase, then starts slowly growing again due to the contraction. The bottom right panel shows that, at a fixed reference time after the conversion, the following approximate scaling holds:
\begin{equation} \label{eq:scalingSqueezing}
\Delta \varphi \propto k, \qquad k \lesssim 10^{-6}  \;.
\end{equation}
Moreover, our numerical results show that, for sufficiently small $k$, the classical value of the temporal phase after the conversion phase is much smaller than its variance,
\begin{equation}
\frac{ \Delta \varphi}{ \varphi} \sim 10^3,
\end{equation}
but this relation is independent of $k$ for sufficiently small $k.$ Thus, the classical value of the angle $\varphi$ also scales in proportion to $k.$ Hence the classical relation between $ v$ and $ \pi$ is essentially $ \pi = 0$ and the inclination of the Wigner ellipse is effectively invisible for small-$k$ modes (see Fig. \ref{FigSqueezing}, top right panel). This implies that for all long-wavelength modes, the initial temporal phase is zero. If one describes the acoustic oscillations of the density perturbations upon horizon re-entry as a linear sum of a $\cos$ and a $\sin$ solution, then our results imply that (for all observationally relevant scales) purely the $\cos$ mode is realized, and consequently all modes with the same wavenumber $k$ will reach maximal and minimal amplitudes in synchrony. This is precisely what is needed to reproduce the acoustic oscillations observed in the cosmic microwave background \cite{Dodelson:2003ip}.

Let us briefly contrast our results with the two-field inflationary model studied in \cite{Prokopec:2006fc}, where the temporal phase was found to have a definite classical value $ \frac{ \Delta \varphi}{ \varphi}  \sim 10^{-3},$ where this ratio is much smaller than in our case. However, it should be clear now that the precise numerical value of this ratio is rather unimportant -- what really matters is that for all observationally relevant scales both the numerator and the denominator in this expression become very small. And this occurs both in the inflationary model of \cite{Prokopec:2006fc} and in the model that we study here.

\begin{figure}[t!]
\begin{minipage}{\smallWidthLeft}
\flushleft
\includegraphics[width=\smallWidthRight]{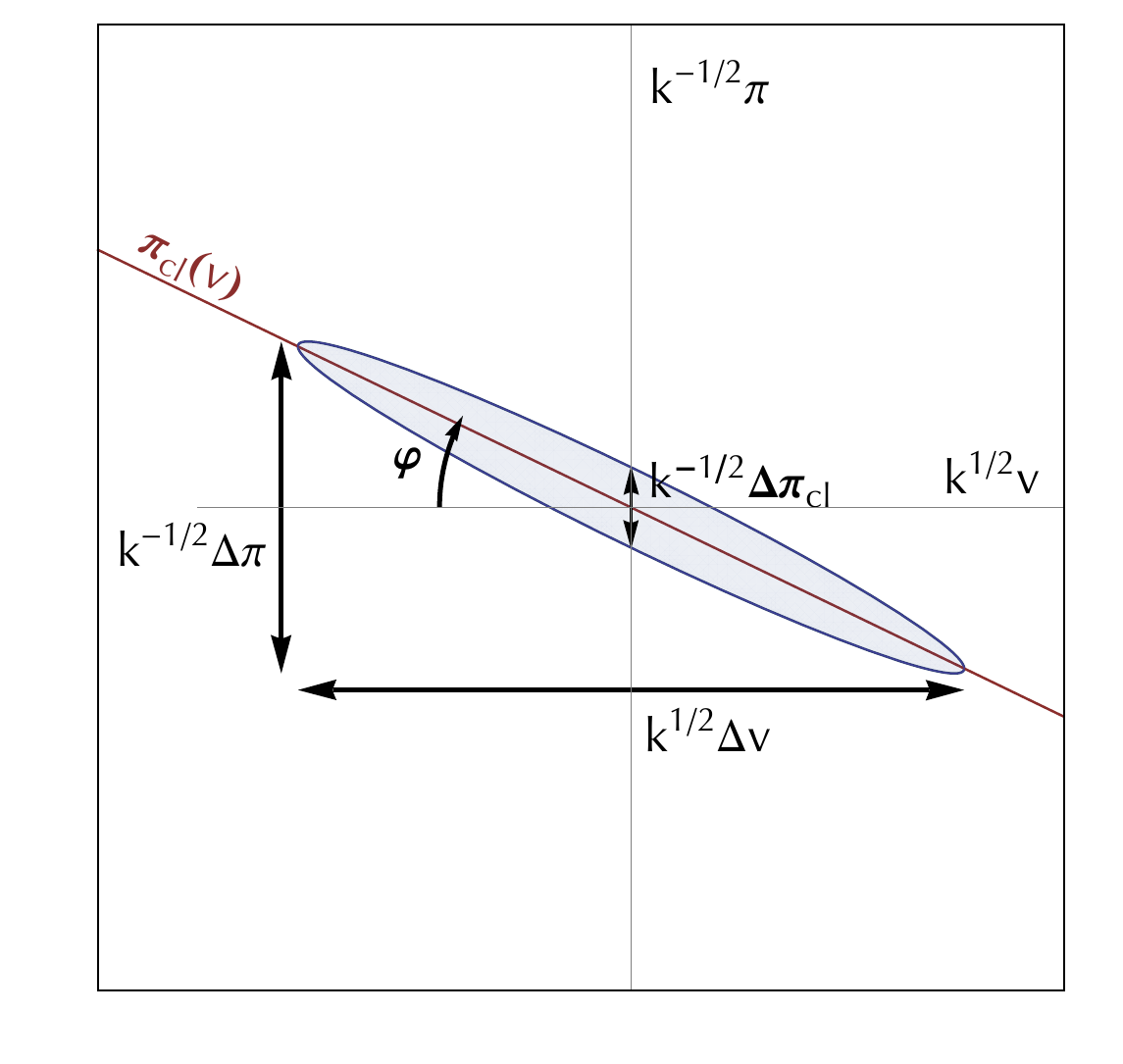}
\end{minipage}%
\begin{minipage}{\smallWidthRight}
\flushleft
\includegraphics[width=\smallWidthRight]{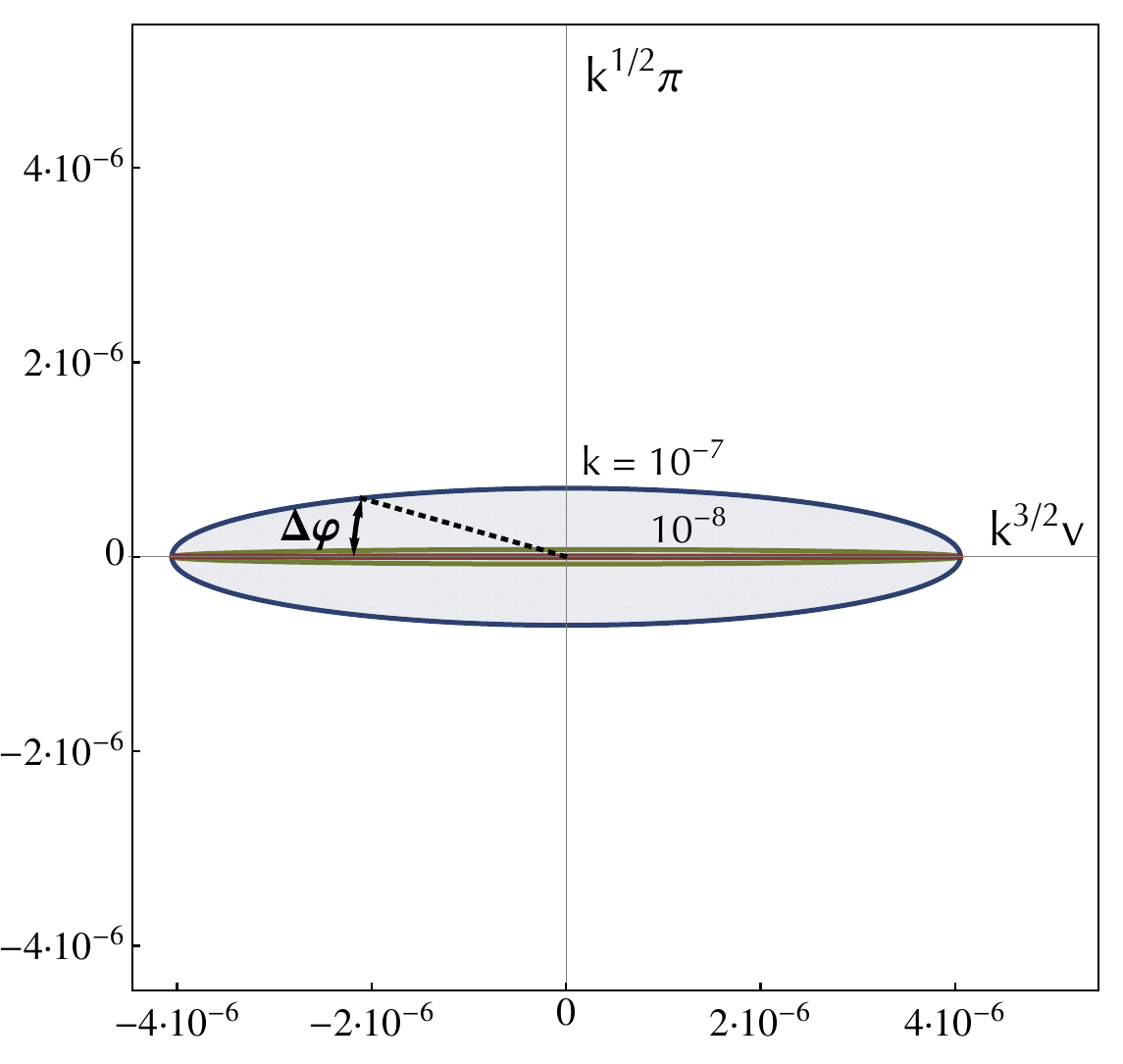}
\end{minipage} \vspace{1cm}\\
\begin{minipage}{\smallWidthLeft}
\flushleft
\includegraphics[width=\smallWidthRight]{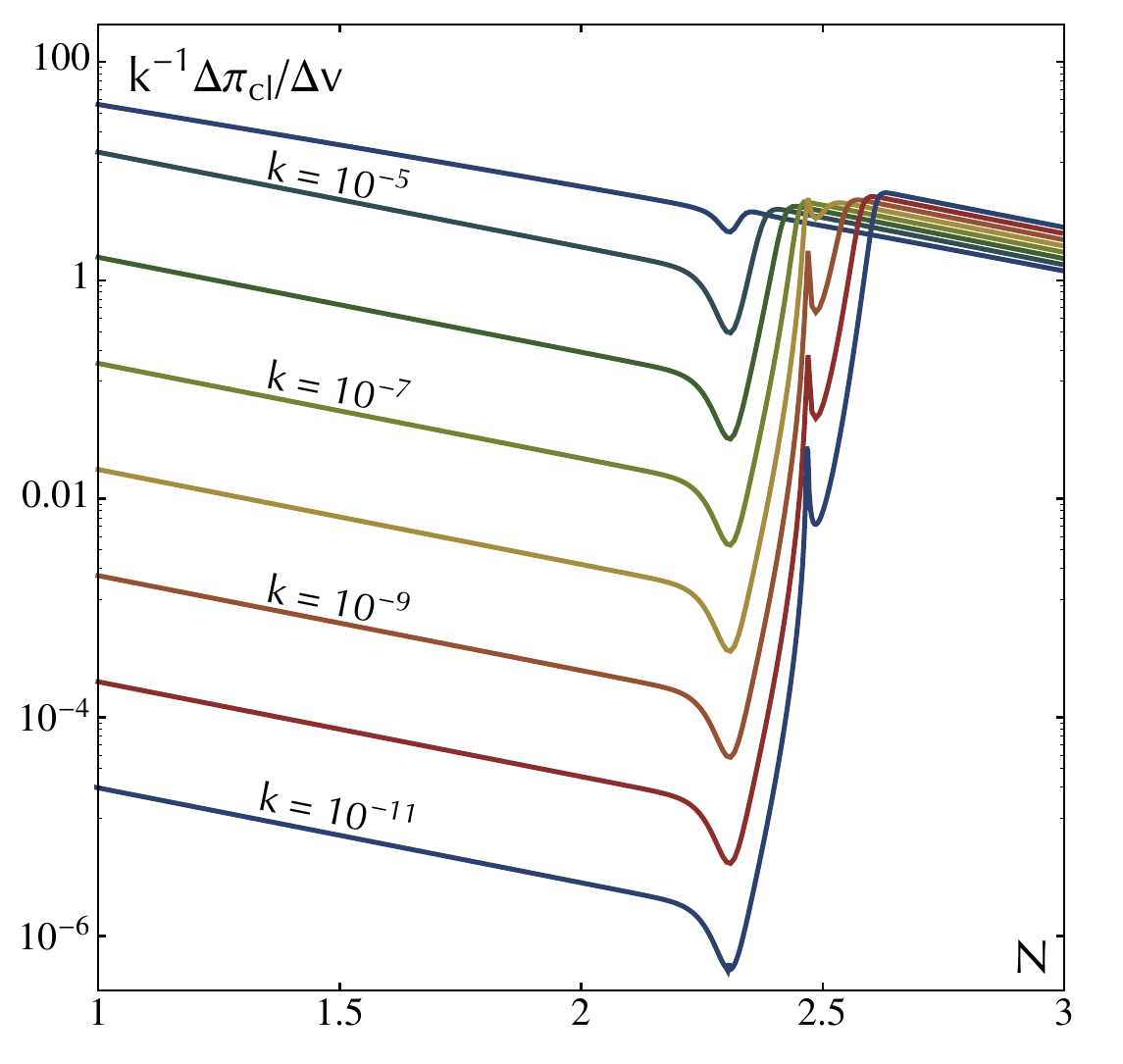}
\end{minipage}%
\begin{minipage}{\smallWidthRight}
\flushleft
\includegraphics[width=\smallWidthRight]{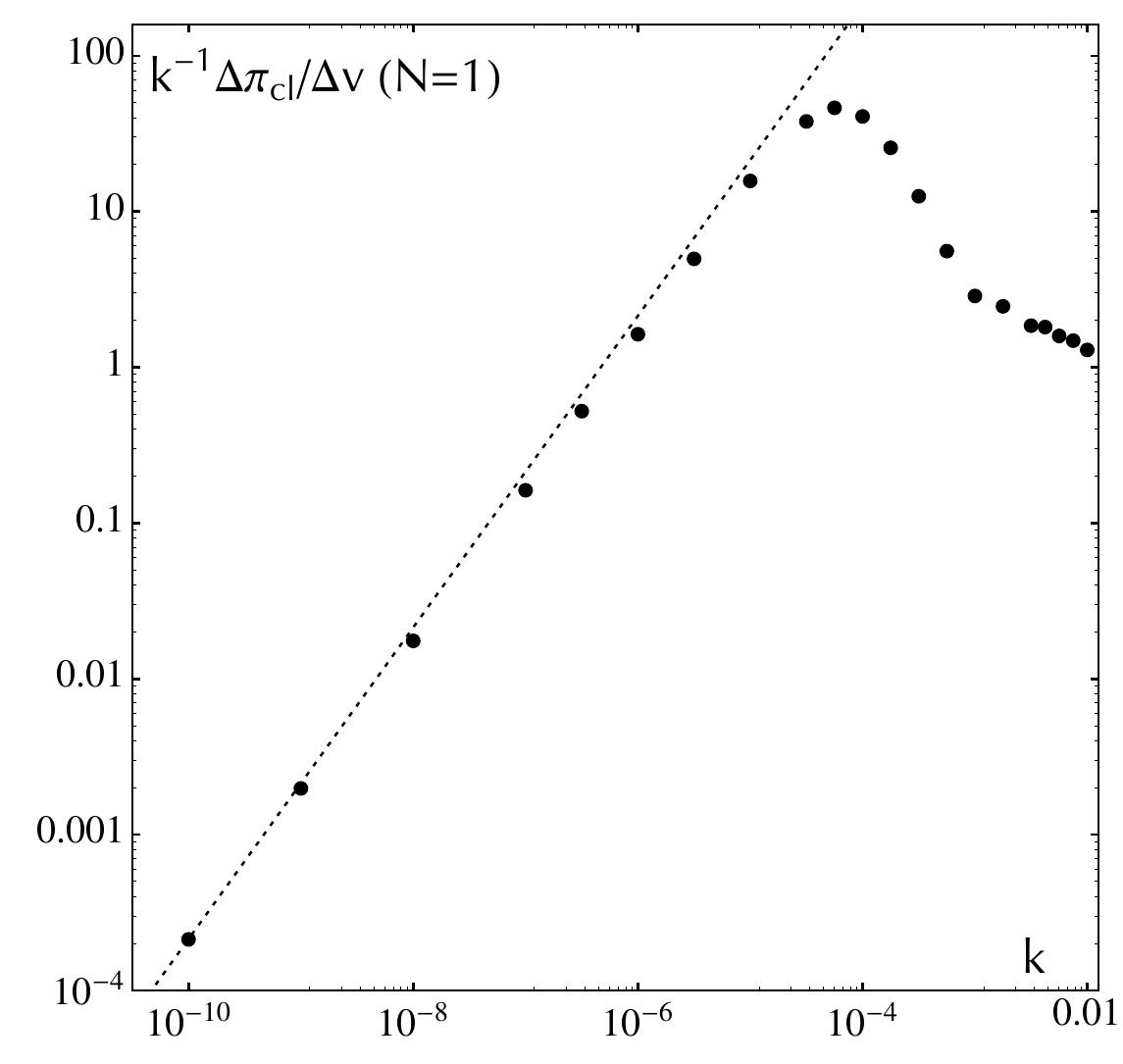}
\end{minipage}\vspace{.7cm}
\caption{ \small \label{FigSqueezing} Top, left panel: general shape of the Wigner ellipse for a gaussian state. Top, right panel: shape of the Wigner ellipses in the rescaled phase space at $N=1$ for $k=10^{-7}$,$10^{-8}$ and $10^{-9}$. The plot shows that the dispersion of $v$ scales very accurately as $k ^{-3/2}$, while the dispersion of the temporal phase rapidly goes to zero: the $k=10^{-9}$ ellipse is barely distinguishable from a segment of the $ \pi= 0$ line. Bottom panels: behavior of the temporal phase dispersion \eqref{eq:deltaPhik} as a function of $N$ and at $N=1$ for different values of $k$. The right panel indicates that $ \Delta \varphi \propto k $ (dotted line) for sufficiently small $k$.
}
\end{figure}

A final comment: the scaling \eqref{eq:scalingSqueezing} can be seen as a consequence of the approximate conservation, at the quantum level, of the comoving curvature perturbation
\begin{equation}
\hat{\mathcal{R} } \equiv \frac{ \hat{v} _{\sigma}}{ z} \;.
\end{equation}
The Heisenberg equation for $ \hat{\mathcal{R} }$ reads (see \eqref{Canonicalmomenta})
\begin{equation}
z \hat{\mathcal{R} }' = \hat{v} _{\sigma}' - \frac{z'}{z} \hat{v} _{\sigma} = \hat{ \pi} _{\sigma} + 2 \theta' \hat{ v} _{s} \;.
\end{equation}
Therefore, after the conversion $( \theta ' = 0 )$ we have approximately
\begin{equation}
z \hat{\mathcal{R} }'  = \hat{ \pi} _{\sigma} \;.
\end{equation}
Hence, the degree of conservation of $ \mathcal{R}$ is directly connected to the dispersion of $ \pi_ \sigma$. In our case, this quantity is simply related to the dispersion of the temporal phase, since the Wigner ellipse is almost horizontal and $ \Delta \pi _{cl} \simeq \Delta \pi$:
\begin{equation}
\Delta \varphi \simeq \frac{ \Delta \pi}{ k \Delta v} \;.
\end{equation}
As previously shown, after the conversion $ \Delta v \propto k ^{-3/2}$, hence
\begin{equation}
\Delta \mathcal{R}' \equiv \langle  \hat{\mathcal{R} } ^{\prime 2} \rangle^{1/2} \propto k ^{1/2} \;.
\end{equation}
This confirms that, after the conversion phase, long wavelength adiabatic curvature perturbations evolve classically in the sense that they are very accurately conserved at the full quantum level.

\section{Discussion}

The proposition put forward both by the inflationary theory of the early universe and by the alternative ekpyrotic/cyclic models is that all structure in the universe originated out of primordial quantum fluctuations, generated either during the currently expanding phase of the universe respectively in a prior contracting phase. This stunning proposition requires that there was a phase in the history of the universe when the usually tiny quantum fluctuations were amplified in such a way as to end up behaving as classical density perturbations.

In the context of inflationary cosmology, it became progressively clear over the last three decades that the quantum-to-classical transition of perturbations relies on several key ingredients: for one, the approximately Gaussian state of the perturbations becomes amplified and highly squeezed during the inflationary phase, as the fluctuation modes exit the horizon. Secondly, decoherence must occur so that one can explain why these squeezed states can be interpreted as a classical ensemble of density perturbations. A final, still unresolved aspect, is to explain why we observe one particular outcome of this classicalization process on the microwave sky, rather than a different (but statistically very similar) one\footnote{In a many-worlds interpretation of quantum mechanics, this is the question of why we happen to find ourselves in one particular decohered branch of the wavefunction rather than another one.}\cite{Martin:2012pea}.

Here, we have performed an analogous analysis for ekpyrotic models. Because the background dynamics is very different in these models, it was not {\it a priori} clear that a similar quantum-to-classical transition could occur here as well. And in fact, during the ekpyrotic phase, the adiabatic curvature perturbations, which have a blue spectrum, are neither amplified nor do they get squeezed. Hence they cannot be interpreted classically. This result, which was previously noticed in \cite{Tseng:2012qd}, has implications for single-field ekpyrotic models: it was originally thought that such blue modes could end up with a scale-invariant spectrum due to (essentially classical) matching conditions at the bounce \cite{Tolley:2003nx}. This now appears unlikely, as the modes of interest cannot be treated classically in the approach to the bounce.

However, in the entropic mechanism nearly scale-invariant entropy perturbations are created during the ekpyrotic phase. These do get amplified and evolve into a highly squeezed state. The subsequent conversion phase kills two birds with one stone: as the entropy perturbations source the curvature perturbations, the latter perturbations inherit the desirable properties of the entropic modes (amplified, squeezed, nearly scale-invariant), while on top of that the interactions between the two types of fluctuations lead to efficient decoherence of the density matrix. The end result is that the entropic mechanism generates a classical ensemble of nearly scale-invariant curvature perturbations in the approach to the bounce. In terms of generating nearly scale-invariant curvature perturbations during the early history of our universe, our results demonstrate that ekpyrotic models can now be considered as truly standing on the same footing as inflationary models.

Recent studies of non-singular bounce models involving higher-derivative kinetic terms for the scalar fields \cite{Cai:2013vm,Cai:2013kja}, loop quantum cosmology models \cite{Wilson-Ewing:2013bla}, and in particular the recent fully non-perturbative classical study of \cite{Xue2013} all suggest that the perturbations evolve through the bounce unscathed, and emerge in the currently expanding phase of the universe in agreement with cosmic microwave background observations. However, it is certainly the case that the bounce phase remains the least understood part of ekpyrotic/cyclic models, and it is here that future progress is most eagerly awaited.

\acknowledgements

We would like to thank Anna Ijjas and Paul Steinhardt for their careful reading of our manuscript, and for very useful suggestions; and we would like to thank Claus Kiefer for an enjoyable and illuminating discussion. We gratefully acknowledge the support of the European Research Council via the Starting Grant Nr. 256994 ``StringCosmOS''.

\appendix

\section{General formalism for two-field cosmological perturbations} \label{section:GeneralFormalism}

In this appendix, we will present the general formalism for quantizing cosmological models with two gauge-invariant scalar perturbation modes as well as the related discussion of the associated reduced density matrix and decoherence. We do this for two reasons: the first is that we hope that this appendix will be useful to the reader when investigating closely related situations. And the second is to highlight the ambiguities that appear when integrating the Lagrangian by parts. Indeed, it is usually assumed that Lagrangians related by total derivatives are equivalent. This is of course true in the sense that they lead to the same equations of motion. However, in quantizing the system, one is necessarily making a choice of canonical variables, and these variables depend on the Lagrangian one is using. In discussing decoherence, we have considered the situation where one of the two fields acts as an environment for the other. Choosing different canonical variables then implies that one is tracing out different degrees of freedom, and the physical results do depend on that choice. Below, we will elaborate more precisely how the results can change by performing such an integration by parts on the original Lagrangian. Before doing so, we should comment on our choice of Lagrangian: we have used the Lagrangian that one obtains directly by substituting the definitions of gauge-invariant perturbations modes, without performing any integrations by parts. This appears to us to be the most conservative and best-motivated choice. However, it would certainly be interesting to try to elucidate further the role of other Lagrangians that are equivalent to ours up to total derivative terms.

We will consider the following general quadratic lagrangian for the real/imaginary part of the Fourier modes of two gauge-invariant perturbation modes $v_\s$ and $v_s$
\begin{eqnarray}\nonumber
L & = & \frac{ \kappa_ \sigma}{2}  v_ \sigma ^{ \prime 2} - \ell_{ \sigma \sigma} v _ \sigma v_ \sigma' - \frac{1}{2} m_{ \sigma} ^2 v_ \sigma ^2 + \frac{ \kappa _{s}}{2}  v_ s ^{ \prime 2} - \ell_{ s s} v _ s v_ s' - \frac{1}{2} m_{s} ^2 v_ s ^2 \\
&&  + \kappa _{ \sigma s} v_ \sigma' v _{s}'-  \ell_{ \sigma s} v_ \sigma' v_s - \ell_{ s \sigma} v_ \sigma v_s' - m_{ \sigma s} ^2 v_ \sigma v_s \;,
\end{eqnarray}
where all the coefficients are \textit{a priori} arbitrary functions of time. Integration by parts maintains the form of the Lagrangian, except for the re--definitions
\begin{eqnarray}
\sigma - \sigma: && \left\{
\begin{array}{rcl}
\ell_{ \sigma \sigma}  & \rightarrow & \lambda_{ \sigma \sigma} \ell _{ \sigma \sigma} \;,\\
m_{ \sigma} ^2 & \rightarrow & m _{ \sigma} ^2 - (1 - \lambda_{ \sigma \sigma}) \ell _{ \sigma \sigma}' \;.
\end{array}
\right. \;, \\
s - s: && \left\{
\begin{array}{rcl}
\ell_{ s s}  & \rightarrow & \lambda_{ s s} \ell _{ s s} \;,\\
m_{ s} ^2 & \rightarrow & m _{ s} ^2 - (1 - \lambda_{ s s}) \ell _{ s s}' \;.
\end{array}
\right.\\
\sigma - s: && \left\{
\begin{array}{rcl}
\ell_{ \sigma s}  & \rightarrow & \lambda_{ \sigma s} \ell _{ \sigma s} \;,\\
\ell_{ s \sigma} & \rightarrow & \ell_{ s \sigma} - (1 - \lambda_{ \sigma s}) \ell_{ \sigma s} \;,\\
m_{ \sigma s} ^2 & \rightarrow & m_{ \sigma s} ^2 - (1 - \lambda_{ \sigma s}) \ell_{ \sigma s}' \end{array}
\right.\\
s- \sigma : && \left\{
\begin{array}{rcl}
\ell_{ \sigma s} & \rightarrow & \ell_{ \sigma s} - (1 - \lambda_{ s \sigma}) \ell_{ s \sigma} \;,\\
\ell_{ s \sigma}  & \rightarrow & \lambda_{ s \sigma} \ell _{ s \sigma} \;,\\
m_{ \sigma s} ^2 & \rightarrow & m_{ \sigma s} ^2 - (1 - \lambda_{ s \sigma}) \ell_{ s \sigma}' \end{array}
\right.
\end{eqnarray}
where the functions $\lambda_{\s\s},\lambda_{\s s},\lambda_{ss}$ depend on the specific integration by parts one is performing. All physical quantities depend only on the invariant combinations
\begin{eqnarray} \label{eq:inv1}
\bar{ m} _{\sigma} ^2 & \equiv & m _{ \sigma} ^2 - \ell_{ \sigma \sigma}' \;,\\  \label{eq:inv2}
\bar{m} _{s}^2 & \equiv &m _{ s} ^2 - \ell_{ s s}' \;,\\  \label{eq:inv3}
\ell & \equiv & \ell _{ \sigma s} - \ell _{ s \sigma} \;,\\  \label{eq:inv4}
\bar{m} _{ \sigma s} ^2 & \equiv & m _{ \sigma s} ^2 - \ell _{ \sigma s}' = m _{ \sigma s} ^2 - \ell _{ s \sigma}' - \ell' \;.
\end{eqnarray}
For instance,  the Euler--Lagrange equations read
\begin{eqnarray}
\left( \kappa_ \sigma v_ \sigma' + \kappa _{ \sigma s} v _{s}' \right)' & = &  - \bar{ m} _{ \sigma} ^2  v_ \sigma + \ell\, v_s' - \bar{ m}_{ \sigma s} ^2 v_ s \;, \\
\left( \kappa_ s v_ s' + \kappa _{ \sigma s} v _{ \sigma}' \right)' & = &  - \bar{ m} _{s} ^2 v_ s - \ell\, v_ \sigma' -  \bar{ m} _{ \sigma s} ^2 v_ \sigma \;.
\end{eqnarray}
However, adding a total derivative to the Lagrangian corresponds to performing a canonical transformation acting on momenta in the Hamiltonian formalism according to
\begin{eqnarray} \label{eq:canonical1}
\sigma - \sigma: && \left\{
\begin{array}{rcl}
\pi_ \sigma  & \rightarrow & \pi_ \sigma + ( 1 - \lambda_ { \sigma \sigma}) \ell_{ \sigma \sigma} v_ \sigma \\
\pi_ s  & \rightarrow & \pi _{s}
\end{array}
\right. \;, \\ \label{eq:canonical2}
s - s: && \left\{
\begin{array}{rcl}
\pi_ \sigma  & \rightarrow & \pi _{ \sigma} \\
\pi_ s  & \rightarrow & \pi_s + ( 1 - \lambda_ { ss}) \ell_{ ss} v_s
\end{array}
\right. \;, \\ \label{eq:canonical3}
\sigma - s: && \left\{
\begin{array}{rcl}
\pi_ \sigma  & \rightarrow & \pi _{ \sigma} + ( 1 - \lambda_{ \sigma s}) \ell_{ \sigma s} v_s \\
\pi_ s  & \rightarrow & \pi_s + ( 1 - \lambda_{ \sigma s}) \ell_{ \sigma s} v_ \sigma
\end{array}
\right. \;, \\ \label{eq:canonical4}
s- \sigma: && \left\{
\begin{array}{rcl}
\pi_ \sigma  & \rightarrow & \pi _{ \sigma} + ( 1 - \lambda_{ s\sigma}) \ell_{ s \sigma} v_s \\
\pi_ s  & \rightarrow & \pi_s + ( 1 - \lambda_{ s \sigma }) \ell_{ s\sigma } v_ \sigma
\end{array}
\right. \;.
\end{eqnarray}
We note that each of these transformations can be written as $\mathbf{ \pi} \rightarrow \pi + M\, v ,$
where $M$ is a \textit{symmetric} matrix.

Proceeding to the quantization of these modes, we write the general solution of the Heisenberg operator equations as
\begin{equation}
\hat{ \mathbf{v}} = \mathbf{F}\hat{a} + \mathbf{F} ^{*}\hat{a} ^{\dagger} + \mathbf{G} \hat{b} + \mathbf{G} ^{*} \hat{b} ^{\dagger} \;,
\end{equation}
where $ \mathbf{F}$ and $ \mathbf{G}$ are complex, linearly independent solutions of the equations of motion and $\hat{a}, \hat{a} ^{\dagger}, \hat{b}, \hat{b} ^{\dagger}$ are two pairs of annihilation/creation operators. Due to the symmetry of the matrix $M,$ the following Wronskian quantities are constants of the motion, whose values we fix in the canonical fashion:
\begin{eqnarray}
\mathbf{F} \cdot \mathbf{P}_F ^{*} - \mathbf{F} ^{*} \cdot \mathbf{P}_F = 1 \;,\\
\mathbf{G} \cdot \mathbf{P}_G ^{*} - \mathbf{G} ^{*} \cdot \mathbf{P}_G  = 1\;,\\
\mathbf{F} \cdot \mathbf{P}_G - \mathbf{P}_F \cdot \mathbf{G} = 0 \;,\\
\mathbf{F} \cdot \mathbf{P}_G ^{*} - \mathbf{P}_F \cdot \mathbf{G} ^{*} = 0 \;.
\end{eqnarray}
Here the quantities denoted $\mathbf{P}_{F,G}$ are the canonical momenta associated with $\mathbf{F},\mathbf{G}.$ One can then re--express the annihilation operators as
\begin{eqnarray}
i\, \hat{a} & = &\mathbf{P}_F ^{*} \cdot \hat{ \mathbf{v}} - \mathbf{F} ^{*} \cdot \hat{ \pi}  \;,\\
i\,\hat{b} & = & \mathbf{P}_G ^{*} \cdot \hat{ \mathbf{v}} - \mathbf{G} ^{*} \cdot \hat{ \pi} \;.
\end{eqnarray}
These expressions can be used to define a vacuum state
\begin{eqnarray}
\hat{a} \ket{0} = \hat{b} \ket{0} = 0 \;,
\end{eqnarray}
which will depend on the choice of mode functions. The vacuum can equally well be described by the wave--function
\begin{eqnarray}
\Psi_{\ket{0}} & \propto &  \textrm{exp} \left( - \frac{1}{2}A_{ \sigma \sigma} v_ \sigma ^2 - A_{ \sigma s} v_ \sigma v_ s - \frac{1}{2} A_{ s s} v_ s ^2 \right) \;,
\end{eqnarray}
\begin{eqnarray} \label{eq:solCorr1}
A_{ \sigma \sigma} & = & - i \frac{( \kappa _{\sigma}F_ \sigma ^{ \prime *} + \kappa _{\sigma s} F _{s} ^{ \prime *})G _{s} ^{*} - F _{s} ^{*} ( \kappa _{\sigma} G _{\sigma} ^{ \prime *} + \kappa _{\sigma s} G _{s}^{ \prime *})}{F _{\sigma} ^{*} G _{s} ^{*} - F _{s} ^{*} G _{\sigma} ^{*}} + i \ell_{ \sigma \sigma} \;,\\ \label{eq:solCorr2}
A_{ ss} & = & - i \frac{( \kappa _{s} F_ s ^{ \prime *} + \kappa _{\sigma s} F _{\sigma} ^{ \prime *}) G _{ \sigma} ^{*} - F _{ \sigma} ^{*} ( \kappa _{s} G _{s}^{ \prime *} +  \kappa _{\sigma s} G _{\sigma} ^{ \prime *})}{F _{s} ^{*} G _{ \sigma} ^{*} - F _{ \sigma} ^{*} G _{ s} ^{*}} + i \ell_{ s s} \;, \\ \label{eq:solCorr3}
A_{ \sigma s} & = & -i \frac{ ( \kappa _{\sigma} F _{\sigma} ^{ \prime *} + \kappa _{\sigma s} F _{s}^{ \prime *}) G _{\sigma} ^{*} - F _{\sigma} ^{*} ( \kappa _{\sigma} G _{\sigma} ^{ \prime *} + \kappa _{\sigma s} G _{s} ^{ \prime *})}{F _{s} ^{*} G _{\sigma} ^{*} - F _{\sigma} ^{*} G _{s} ^{*}} + i \ell _{ \sigma s} \\
& = & -i \frac{ ( \kappa _{s} F _{s} ^{ \prime *} + \kappa _{\sigma s} F _{ \sigma}^{ \prime *}) G _{s} ^{*} - F _{s} ^{*} ( \kappa _{ s} G _{ s} ^{ \prime *} + \kappa _{\sigma s} G _{ \sigma} ^{ \prime *})}{F _{ \sigma} ^{*} G _{s} ^{*} - F _{s} ^{*} G _{ \sigma} ^{*}} + i \ell _{ s \sigma} \;.
\end{eqnarray}
Not surprisingly, the correlators depend on the quantization: the different behaviors of the phase of the wave--function correspond to different values of momenta, related by (\ref{eq:canonical1}--\ref{eq:canonical4}). By considering the time--dependent Schr\"odinger equation, we can obtain equations of motion for the correlators. For this, we need an expression for the Hamiltonian of our system. It is given by
\begin{eqnarray} \nonumber
H & = & \frac{ \kappa_s}{2 \kappa} \left( \pi _{\sigma} + \ell  _{\sigma\sigma} v _{\sigma} + \ell _{\sigma s} v _{s} \right) ^2 + \frac{ \kappa_ \sigma}{2 \kappa} \left( \pi _{s} + \ell  _{s} v _{s} + \ell _{s \sigma} v _{ \sigma} \right) ^2 - \\ \nonumber
&&-\frac{ \kappa_{ \sigma s}}{ \kappa} \left( \pi _{\sigma} + \ell  _{\sigma\sigma} v _{\sigma} + \ell _{\sigma s} v _{s} \right)\left( \pi _{s} + \ell  _{s} v _{s} + \ell _{s \sigma} v _{ \sigma} \right) \\
&& + \frac{1}{2} m _{\sigma} ^2 v _{\sigma} ^2 + \frac{1}{2} m _{s} ^2 v _{s} ^2 + m _{\sigma s} ^2 v _{\sigma} v_{s} \;,\\
\kappa & \equiv & \kappa_ \sigma \kappa _{s} - \kappa _{\sigma s} ^2 \;,
\end{eqnarray}
and it can be quantized trivially since the ordering ambiguities correspond to an additive constant in $H$. The equations for the correlators, obtained from $i \Psi' = \hat{H} \Psi,$ or directly from the equations of motion for the constituent mode functions, take the form
\begin{eqnarray}
-i \kappa \bar{A}_{ \sigma \sigma}' & = & \kappa\,\bar{m} _{\sigma} ^2 - \kappa _{s} \bar{A} _{ \sigma \sigma} ^2 - \kappa _{\sigma} \left( \bar{A} _{ \sigma s} + i \ell\right)  ^2 + 2 \kappa _{\sigma s} \bar{A} _{\sigma\sigma} ( \bar{A} _{\sigma s} + i \ell) \;,\\
- i \kappa \bar{A} _{ ss}' & = & \kappa\, \bar{m} _{s} ^2 - \kappa _{\sigma} \bar{A} _{ ss} ^2 - \kappa _{s} \bar{A} _{ \sigma s} ^2 + 2 \kappa _{\sigma s} \bar{A} _{ss}\bar{A} _{\sigma s} \;,
\\
- i \kappa \bar{A} _{ \sigma s}' & = & \kappa\, \bar{m}_{ \sigma s} ^2 - \left( \kappa _{s} \bar{A} _{ \sigma \sigma} + \kappa _{\sigma} \bar{A}_ { ss} \right) \bar{A} _{ \sigma s} - i \kappa _{\sigma} \ell \bar{A} _{ ss}  + \kappa _{\sigma s} \left( \bar{A} _{\sigma\sigma} \bar{A} _{ss} + \bar{A} _{\sigma s}( \bar{A} _{\sigma s} + i \ell) \right)  \;,\\
\bar{A}_{ \sigma \sigma} & \equiv & A _{ \sigma \sigma} - i \ell _{\sigma\sigma} \;,\\
\bar{A} _{\sigma s} & \equiv & A _{ \sigma s} - i \ell _{\sigma s} \;,\\
\bar{A} _{ss} & \equiv & A _{ss} - i \ell _{ss}\;.
\end{eqnarray}
Note that the equations for the reduced correlators $ \bar{A}$ only contain the invariant combinations (\ref{eq:inv1}-- \ref{eq:inv4}). Moreover, according to (\ref{eq:solCorr1}-- \ref{eq:solCorr3}) the initial conditions for the correlators are also independent of the quantization. Thus, so far all results are unaffected by performing an integration by parts on the Lagrangian.

However, in discussing the reduced density matrix and decoherence, this is no longer so: tracing over {\it e.g.} $ v_s$, the reduced density matrix for measurements of $v_ \sigma$ reads
\begin{eqnarray}
\rho( \bar{v}_ \sigma, v_ \sigma) & = & N'\, \textrm{exp} \left( - \frac{1}{2} C_{SS} v_S ^2 - \frac{1}{2} C_{DD} v_D ^2 - i C_{SD} v_S v_D \right) \;,\\
v_S & \equiv & \frac{1}{2} \left(v_ \sigma + \bar{v} _{\sigma} \right) \;,\\
v_D & \equiv & v_ \sigma - \bar{v} _{\sigma} \;, \\
C_{SS} & = & 2 A _{ \sigma \sigma} ^{R} \left( 1 - \frac{( A _{ \sigma s} ^{R})^2}{A_{ss} ^{R} A _{ \sigma \sigma} ^{R}} \right) \;,\\
C_{SD} & = & A_{ \sigma \sigma} ^{I} \left(1 - \frac{A _{ \sigma s} ^{I} A_{ \sigma s} ^{R}}{A_{ s s} ^{R}A_{ \sigma \sigma} ^{I}} \right) \;, \\ C_{DD} & = & \frac{1}{2} A_{ \sigma \sigma} ^{R} \left( 1 + \frac{( A_ { \sigma s} ^{I}) ^2}{ A_{ss} ^{R} A_{ \sigma \sigma} ^{R}} \right) \;.
\end{eqnarray}
As discussed in the previous sections, the main indicator for decoherence is the ratio
\begin{equation}
\frac{C_{DD}}{C_{SS}} = \frac{\bar{A} ^{R} _{ss} \bar{A} ^{R} _{\sigma\sigma} + ( \bar{A} ^{I} _{\sigma s} + \ell_{ \sigma s}) ^2}{ \bar{A} ^{R} _{ss} \bar{A} ^{R} _{\sigma\sigma} - ( \bar{A} ^{R} _{\sigma s}) ^2} \;.
\end{equation}
This quantity {\it does} depend on the quantization, as is obvious from the appearance of the $l_{\s s}$ term. Therefore, the answer for the amount of decoherence depends explicitly on the choice of canonical variables. As an application, we checked that the prediction for the toy model  studied in \cite{Tseng:2012qd}, namely no decoherence for kinetically coupled scalar fields during an ekpyrotic phase, only holds for the choice of variables adopted in that paper ($\ell_{ \sigma s} = a'/a$). On the other hand, if like in our case the two fields are not coupled from a given time on ($\ell_{ \sigma s} \propto \theta' $), this ambiguity is no longer present.

\section{Formulae involving Bessel functions} \label{section:Bessel}
Approximating cosmological mode functions at late times requires the use of the following asymptotic behaviors of Bessel functions (with $\alpha > 0$)
\begin{eqnarray} \label{eq:asy1}
J_ \alpha(x) & \stackrel{ x \rightarrow 0}{ \sim} & \frac{1}{ \Gamma( \alpha+ 1)} \left( \frac{x}{2} \right) ^{\alpha} + \mathcal{O}( x ^{ \alpha + 2})\;, \\ \label{eq:asy2}
Y_ \alpha(x) & \stackrel{ x \rightarrow 0}{ \sim} & - \frac{ \Gamma( \alpha)}{ \pi} \left( \frac{x}{2} \right) ^{ - \alpha} - \frac{ \Gamma( -\alpha) \cos{( \pi \alpha})}{ \pi} \left( \frac{x}{2} \right) ^{ \alpha} - \frac{ \Gamma( \alpha-1)}{ \pi} \left( \frac{x}{2} \right) ^{2-\alpha} + \mathcal{O}(x ^{ \alpha + 2})\;.
\end{eqnarray}
In order to work out the corresponding correlators, one needs to evaluate time derivatives of the mode functions. Denoting $x \equiv - k \tau,$ so that $v'=-k v_{,x},$ we have
\begin{eqnarray*}
\frac{d v}{dx}   & = & \sqrt{\frac{\pi}{4k}}\frac{1}{ 2 \sqrt{x}} \left( J_ \alpha + 2 x \frac{dJ _{ \alpha}}{dx} + i Y _{ \alpha} + 2 i x \frac{dY _{ \alpha}}{dx}  \right) \;,
\end{eqnarray*}
and hence
\begin{eqnarray*}
\textrm{Re} \left(\frac{ v ^{\prime *}}{ v^{*}} \right)
& = & \frac{1}{2 \tau} - \frac{k}{2 (J _{ \alpha} ^2 + Y _{ \alpha} ^2)} \frac{d}{dx} \left(J _{ \alpha} ^2 + Y _{ \alpha} ^2 \right) \;, \\
\textrm{Im} \left(\frac{ v ^{ \prime *}}{ v ^{*}} \right)
& = & - \frac{k}{J _{ \alpha} ^2 + Y _{ \alpha} ^2} \left[ Y _{ \alpha} \dot{J} _{ \alpha} - J _{ \alpha} \dot{Y} _{ \alpha} \right] \;.
\end{eqnarray*}
It turns out that we must keep sub--leading terms in order to calculate the asymptotic behavior of the correlators. A straightforward calculation leads to
\begin{eqnarray*}
J _{ \alpha} ^2 + Y _{ \alpha} ^2 & = & \frac{ \Gamma( \alpha) ^2}{ \pi ^2} \left( \frac{x}{2} \right) ^{-2 \alpha} \left( 1 + \frac{2 \Gamma(- \alpha) \cos{( \pi \alpha)}}{ \Gamma( \alpha)} \left( \frac{ x}{2} \right) ^{ 2 \alpha} + \frac{2}{ \alpha - 1} \left(\frac{x}{2} \right) ^2 + \ldots \right) \;,\\
\frac{d}{dx} \left(J _{ \alpha} ^2 + Y _{ \alpha} ^2 \right) & = & -\frac{2 \alpha }{x} \left( J _{ \alpha} ^2 + Y _{ \alpha} ^2 \right) \left(1 - \frac{2 \Gamma( - \alpha) \cos{( \pi \alpha)}}{ \Gamma( \alpha)} \left( \frac{ x}{2} \right) ^{ 2 \alpha} - \frac{2 }{ \alpha( \alpha - 1)} \left( \frac{x}{2} \right) ^2 + \ldots\right) \;,
\end{eqnarray*}
so that one obtains
\begin{eqnarray*}
\textrm{Re} \left(\frac{ v^{\prime *}}{ v^{*}} \right)
& = & \left( \frac{1}{2} - \alpha \right) \frac{1}{ \tau} + \frac{2 \alpha \Gamma( - \alpha) \cos{ ( \pi \alpha)}}{ \tau \Gamma( \alpha)} \left( \frac{x}{2} \right) ^{2 \alpha} + \ldots  \\
& = & \frac{(1- 2 \alpha)( \epsilon-1)}{2} \mathcal{H} + \frac{2 \alpha \Gamma( - \alpha) \cos{ ( \pi \alpha)}}{ \tau \Gamma( \alpha)} \left( \frac{x}{2} \right) ^{2 \alpha} + \ldots \;, \\
\textrm{Im} \left(\frac{ v^{ \prime *}}{ v^{*}} \right) & = &  - \frac{ 2 \pi}{ \tau \Gamma( \alpha) ^2} \left( \frac{x}{2} \right) ^{2 \alpha} + \ldots \;.
\end{eqnarray*}

\bibliographystyle{utphys}
\bibliography{masterQC}

\end{document}